\documentclass[pre,aps,twocolumn,floatfix, superscriptaddress,showpacs]{revtex4-1}
\usepackage{amsmath,amssymb}
\usepackage{setspace}
\usepackage{comment}
\usepackage{graphicx}
\usepackage{psfrag}
\usepackage{multirow}
\usepackage{alltt}
\usepackage{color}

\newcommand{\bt}[1]{{\mathbf #1}}
\def\ds{\displaystyle}

\begin{document}

\title{Modulational Instability in Basic Plasma and Geophysical Models}
\author{Brenda Quinn}
\affiliation{School of Mechanical Engineering, Center for Mediterranean Sea Studies, Tel-Aviv University, Israel}
\author{Sergey Nazarenko}
\affiliation{Mathematics Institute, University of Warwick, Gibbet Hill Road, Coventry CV4 7AL, United Kingdom}
\affiliation{Laboratoire SPHYNX, Service de Physique de l'Etat Condense, DSM, IRAMIS, CEA, Saclay, CNRS URA 2464, 91191, Gif-sur-Yvette, France}
\author{Colm Connaughton}
\affiliation{Mathematics Institute, University of Warwick, Gibbet Hill Road, Coventry CV4 7AL, United Kingdom}
\affiliation{Centre for Complexity Science, University of Warwick, Gibbet Hill Road, Coventry CV4 7AL, United Kingdom}
\affiliation{Okinawa Institute of Science and Technology Graduate University, 1919-1 Tancha Onna-son, Okinawa 904-0495, Japan}
\author{Steven Gallagher}
\affiliation{Department of Physics, University of Warwick, Gibbet Hill Road, Coventry CV4 7AL, United Kingdom}
\author{Bogdan Hnat}
\affiliation{Department of Physics, University of Warwick, Gibbet Hill Road, Coventry CV4 7AL, United Kingdom}

\begin{abstract}
This is a review of the theory of the modulational instability in idealised fluid models of strongly magnetised plasmas and reduced models of geophysical fluid dynamics, particularly the role it plays in the formation of zonal flows. The discussion focusses on the Charney-Hasegawa-Mima and Hasegawa-Wakatani models. Particular attention is paid to the wave turbulence - zonal flow feedback loop whereby large scale zonal flows which are initially generated by modulational instability of small-scale drift/Rossby waves tend to subsequently suppress these small scale waves by their shearing action. This negative feedback can result in a dynamic equilibrium in which large scale zonal flows grow by drawing energy from small scale turbulence but suppress the small scale turbulence in the process until a balance is reached. In this regime, the level of small scale turbulence is greatly reduced compared to the level one would observe in the absence of the zonal flows. 
\end{abstract}

\maketitle

\section{Introduction}

Weak, long-wave modulations of a periodic nonlinear wave train can grow exponentially under certain  conditions.  This  is known as the modulational instability (MI) and was first identified by \cite{BenjamnFeir1967} for surface gravity waves on deep water. MI is a ubiquitous instability in nature and is not restricted to hydrodynamics. It also occurs in plasma physics, electrodynamics and nonlinear optics~\citep{Zakharov2009}. MI is relevant to the present collection because it is one mechanism for generating zonal flows (ZF) in magnetically-confined plasmas and in planetary atmospheres and oceans ~\citep{Diamond2005,Onishchenko2008}.


Plasma drift turbulence and geophysical quasigeostrophic (QG) turbulence are often discussed together~\citep{HasegawaMac1979,Horton1994,Diamond2005} because they
share many structural features. {Both systems are
quasi-two-dimensional and anisotropic, exhibit direct and inverse cascades, 
contain a mixture of wave and vortex dynamics and tend to spontaneously form large scale zonal flows. Consequently}, the
basic linear and nonlinear properties of these two systems can be modeled by same
nonlinear PDE, the Charney-Hasegawa-Mima (CHM) equation  ~\citep{Charney1948,HasegawaMima1978}, introduced in section \ref{sec-models}.
CHM is the simplest useful nonlinear models describing both Rossby waves in the atmosphere and also drift waves in a magnetically-confined plasmax.  Despite its simplicity, it exhibits remarkable behaviour, and many conceptual ideas about the interactions between turbulence and ZFs were conceived within CHM. These ideas provide qualitative understanding of the dynamics in fusion devices, most notably the Low-to-High (LH) confinement transition. In geophysical flows, similar ideas may be traced back to the Barotropic Governor regulating mechanism put forward by \cite{James1987}.

Beyond CHM, in both plasma physics and geophysical fluid dynamics, a hierarchy of models exists with increasing degrees of realism, achieved at a cost of increasing complexity. In plasmas, the next level model is the Extended Hasegawa-Mima (EHM) model that improves the description of the electron response~\citep{Dorland1990} followed by the Hasegawa--Wakatani (HW) model which incorporates the drift-dissipative instability \citep{Hasegawa1983}.  In GFD the next level of description
is the two-layer model which incorporates baroclinic instability~\citep{McWilliams2006}.  Here, we review the MI in the basic nonlinear models mentioned above. We will also briefly review studies of the systems forced by a
primary instability, such as the forced-dissipated CHM and the HW models. The two-layer GFD model is not covered by the present review. In these systems, MI appears at the first evolutionary stage, followed by formation of ZF which feed back on the scales of the the primary forcing and suppress the latter. This is the most basic mechanism containing all the essential features of the LH transitions.  

\section{Basic Nonlinear Models}
\label{sec-models}

\subsection{Charney-Hasegawa-Mima model}

Until the late 1940s numerical calculations of the atmospheric equations of motion remained unduly complicated since they required integrating over the whole range of scales.  It had been pointed out by meteorologists that the motions which contribute to the large scale weather patterns could be classified as quasi-hydrostatic, quasi-adiabatic, quasi-horizontal and quasi-geostrophic~\citep{Charney1948,Lynch2003}.  Using these four approximations, Charney filtered out the noise from the governing fluid equations to leave only the important long wavelength waves that contribute to the large scale weather phenomena in the midlatitudes~\citep{Charney1948,Pedlosky1987}.
 
The HM equation~\citep{HasegawaMima1978} is the simplest single-fluid description of a strongly magnetised plasma and is derived from first principles, beginning with the continuity of ions and allowing a coupling between flux-surface-averaged potentials and density fluctuations.  The densities of ions and electrons are assumed to have a Boltzmann distribution and are adiabatic due to a slow variation of the potential in the parallel direction. 

 Equivalence of  the Charney and Hasegawa-Mima equations was first pointed out by \cite{HasegawaMac1979}. See also \cite{Horton1994}.  It is now usually known as the CHM equation. The equation is
\begin{equation}
\label{eq-Charney}
\frac{\partial}{\partial t}(\nabla^2\psi-F\psi) + \beta \frac{\partial \psi}{\partial x} +\frac{\partial \psi}{\partial x} \frac{\partial \nabla^2\psi}{\partial y} - \frac{\partial \psi}{\partial y}\frac{\partial \nabla^2\psi}{\partial x}  = 0\,,
\end{equation}
where within the plasma context $\psi=\psi(x,y,t)$ is the electrostatic potential, $\ds F=\frac{1}{\rho^2}$ with $\rho$ being the ion Larmor radius at the electron temperature and $\beta$ is a constant proportional to the mean plasma density gradient. { In fusion plasmas, $\rho$ is usually a small scale quantity.} Correspondingly in GFD these variables are the streamfunction {(geopotential height)}, Rossby deformation radius and a constant proportional to the latitudinal gradient of the vertical rotation frequency respectively. { In geophysical turbulence, $\rho$ can be a small scale quantity (for example in the Earth's oceans or in Jupiter's atmosphere) or a large scale quantity (for example in the Earth's atmosphere).} In plasmas (GFD), the $y$-axis is in the radial (south--north) direction, along the plasma density
gradient and the $x$-axis is the poloidal (west--east) direction.  Note that here we are using a GFD convention; in plasma literature the $x$- and the $y$-axes are usually exchanged. Detailed derivations can be found in many texts~\citep{Charney1948,HasegawaMima1978,Pedlosky1987}.

We consider for simplicity a system in a doubly periodic box of size $L$ and take the spatial Fourier transform of the streamfunction, $\hat{\psi}_{\bf{k}} = \int \psi({\bt x}) \mathrm{e}^{-\mathrm{i} (\bf{k} \cdot \bf{x})} \, d \bf{x}$.
Since $\psi_{\bf{x}}$ is real, $\psi_{-\bt{k}} = \overline{\psi}_\bt{k}$, where the overbar denotes the complex conjugate.
In Fourier space Eq.~(\ref{eq-Charney}) is
\begin{equation}
\label{eq-CHMk} 
\partial_t \hat{\psi}_\bt{k} = - i\, \omega_\bt{k}\, \hat{\psi}_\bt{k}  
 + \frac{1}{2} \sum_{\bt{k}_1, \bt{k}_2} T(\bt{k},\bt{k}_1,\bt{k}_2)\, \hat{\psi}_{\bt{k}_1}\, \hat{\psi}_{\bt{k}_2}\, \delta( {\bt{k} - \bt{k}_1 - \bt{k}_2})\,
\end{equation}
where
\begin{equation}
\label{eq-Rossbydispersion}
\omega_\bt{k} = -\frac{\beta k_x}{k^2 + F}
\end{equation}
is the anisotropic dispersion relation for the linear wave solutions, $\bt{k}=(k_x,k_y)= (2\pi \bt{m}/L, 2\pi \bt{n}/L)$ is the wave vector ($\bt{m}, \bt{n} \in Z^2$), $k = \left| \bt{k} \right|$ and
\begin{equation}
\label{eq-CHMinteractioncoeff}
T(\bt{k},\bt{k}_1,\bt{k}_2) = \frac{\left(\bt{k}_1 \times \bt{k}_2\right)_z (k_2^2-k_1^2)}{k^2 + F}
\end{equation}
is the nonlinear interaction coefficient which vanishes for a monochromatic wave. Hence monochromatic Rossby waves and drift waves are exact solutions of the full nonlinear CHM equation. 

\subsection{Extended Hasegawa-Mima model}
In the plasma context, modes with $k_x=0$ must be special because for these
modes the relation between the plasma potential and the density fluctuations (so called Boltzmann response) fails {\citep{Dorland1990,dorlandAndHammett1993}}. In fact, for such modes the density and the potential fields decouple. This effect is taken into account in 
 the extended {(or modified)} Hasegawa-Mima (EHM) equation {(\cite{dorlandAndHammett1993,SmolyakovShev2000,Dewar2007}) in which the coupling between the flux surface averaged potentials and density fluctuations is removed. The equation is :
 \begin{equation}
 \label{eq-EHM-realSpace}
\frac{\partial}{\partial t}(\nabla^2\psi-F(\psi-\delta_{s,1}\bar{\psi})) + \beta \frac{\partial \psi}{\partial x} +\frac{\partial \psi}{\partial x} \frac{\partial \nabla^2\psi}{\partial y} - \frac{\partial \psi}{\partial y}\frac{\partial \nabla^2\psi}{\partial x}  = 0\,,
\end{equation}
where the notation $\bar{f}$ denotes the zonal average of a function, $f$,
\begin{equation}
\bar{f}(y,t) = \frac{1}{2\pi}\int_0^{2\pi} f(x,y,t)\,dx
\end{equation}
and $\delta_{x,y}$ is the Kronecker symbol.}
Following \cite{Dewar2007} we include	 a switch parameter, $s$. Taking $s=1$ gives the EHM and $s=0$ gives the CHM. In Fourier space, Eq.~(\ref{eq-EHM-realSpace}) has the same form as Eq.~(\ref{eq-CHMk}) with $\omega_{\bt{k}}$ still given by Eq.~(\ref{eq-Rossbydispersion}) but 
the interaction coefficient  is modified 
\begin{equation}
\label{eq-EHMinteractioncoeff}
T (\bt{k}, \bt{k_1},\bt{k_2}) = \frac{\left(\bt{k}_1 \times \bt{k}_2\right)_z \left[\rho^2(k_2^2-k_1^2)  + \delta_{s, 1} (\delta_{\bt{k}_{2_x}, 0} - \delta_{\bt{k}_{1_x}, 0})\right]}{(1 +
  \rho^2 k^2 - \delta_{\bt{k}_x, 0} \delta_{s, 1})}\,.
\end{equation}
 The ``extension" of the model (terms involving $\delta$) act only on modes with a $k_x=0$ component. This increases coupling to the zonal modes { and enhances the growth of zonal flows in the EHM model relative to the CHM model.}
Note that the difference between CHM and EHM disappears as $\rho \to \infty$.

\subsection{Forced-dissipated CHM and EHM models} 
Neither the CHM nor the  EHM models can spontaneously generate waves. This shortcoming is fixed in more complex two-field models, namely the Hasegawa-Wakatani model in plasmas and the two-layer QG model in GFD. These models contain forcing by primary instability mechanisms, the drift-dissipative and the baroclinic instabilities respectively. However, one could try to mimic such instabilities by simply adding to the one-field CHM or EHM models extra linear terms which produce the same $\bf k$-space distribution of the linear growth and dissipation rates as the ones predicted by the two-field models.  This is done by modifying the dispersion relation in the respective CHM and EHM models, namely taking Eq.~(\ref{eq-Charney}) with
$\omega_\bt{k}$
 given by
\begin{equation}
\label{eq-driftdispersion}
\omega_\bt{k} = \frac{-\beta k_x}{1+\rho^2 k^2 }
 +i \lambda_{\bt{k}}\,.
\end{equation}
Relative to the unforced model, the additional imaginary part produces instability type forcing ($\lambda_{\bt{k}} >0$) and dissipation ($\lambda_{\bt{k}} <0$).

\subsection{Hasegawa-Wakatani model and extended version}

The Hasegawa-Wakatani (HW) model {\citep{Hasegawa1983}} is more realistic and physical than the CHM since it spontaneously generates waves at the level of linear dynamics via an intrinsic instability. It relaxes the adiabatic relationship between the density and potential and instead assumes that the density response is coupled to the potential via electron dynamics in the direction parallel to the magnetic field. { Instability can therefore arise by the conversion of potential energy stored in density gradients into kinetic energy stored in fluctuations of the electric potential. The  HW model is a pair of coupled equations for the density, $n(x,y,t)$, and potential $\psi(x,y,t)$:  
\begin{eqnarray}
\label{eq-HW-realSpace}
\frac{\partial}{\partial t}(\nabla^2\psi) +\frac{\partial \psi}{\partial x} \frac{\partial \nabla^2\psi}{\partial y} - \frac{\partial \psi}{\partial y}\frac{\partial \nabla^2\psi}{\partial x}  &=& \alpha\,\left(\psi-n\right)\\
\nonumber \frac{\partial n}{\partial t} + \kappa \frac{\partial n}{\partial x} +\frac{\partial \psi}{\partial x} \frac{\partial n}{\partial y} - \frac{\partial \psi}{\partial y}\frac{\partial n}{\partial x} &=&\alpha\,\left(\psi-n\right).
\end{eqnarray}
Here $\alpha$ is a coupling parameter and $\kappa$ is the mean density gradient.
An extended version of the HW model was introduced by \cite{Numata2007}. In Fourier space, the original and extended HW models can be written:}
\begin{eqnarray}
\nonumber \partial_t \hat{\psi}_\bt{k} &=& - \frac{\alpha z}{k^2} (\hat{\psi}_\bt{k} -
\hat{n}_\bt{k})  +
 \frac 1{2} \sum_{\bt{k}_1, \bt{k}_2} T(\bt{k}, \bt{k_1},\bt{k_2}) \hat{\psi}_\bt{k_1}
\hat{\psi}_\bt{k_2}\\
\nonumber &&\times \delta( {\bt{k} - \bt{k}_1 - \bt{k}_2})
\nonumber
\\
\nonumber\partial_t \hat{n}_\bt{k} &=& - i \kappa k_y \hat{\psi}_\bt{k} + z \alpha 
 (\hat{\psi}_\bt{k} - \hat{n}_\bt{k}) \!  -\!\!
\sum_{\bt{k}_1, \bt{k}_2} \!\! R(\bt{k}_1, \bt{k}_2)\\
&&  \times \hat{n}_\bt{k_1}   \hat{\psi}_\bt{k_2} 
\delta( {\bt{k} - \bt{k}_1 - \bt{k}_2})
\label{NonLocalHWChap-HW-N}
\end{eqnarray}
where $z = 1- \delta_{s,1}\delta_{k_y,0}$
and s is the switching parameter as before: s=0 gives the HW case and s=1 gives the EHW case. 
The interaction coefficients are
\begin{eqnarray*}
R (\bt{k_1},\bt{k_2}) &=& \left(\bt{k}_1 \times \bt{k}_2\right)_z\,,\\
\label{eq-HWinteractioncoeff}
T (\bt{k}, \bt{k_1},\bt{k_2}) & =& \frac{\left(\bt{k}_1 \times \bt{k}_2\right)_z (k_2^2-k_1^2)  }{
   k^2 }\,.
\end{eqnarray*}

In the limit $\alpha \gg 1$ the HW/EHW system tends to the familiar CHM/EHM system, whereas in the limit $\alpha \ll 1$
it becomes the 2D Euler equation for the streamfunction $\psi$  and the passive advection
equation for $n$.
Note that for the HW/EHW we have exchanged the $x$-and $y$-axes with respect to our CHM/EHM notations. This is because HW/EHW are purely plasma models and we would like to use the conventions of the plasma literature. (Since CHM is a common model for GFD and plasmas, and since it was first introduced in GFD, we have used geophysical conventions for that model).

\section{Modulational instability in CHM and EHM}
\label{sec-MI}
Given that Eq.~(\ref{eq-CHMk}) admits an exact solution in the form of a monochromatic wave with an arbitrary amplitude,
it is natural to ask how stable these waves are to small perturbations.  {For the moment we consider stability in terms of an initial value problem and defer the question of how such waves are generated in the first place until later.}
Such a problem was first formulated and analysed by
~\cite{Lorenz1972} and ~\cite{Gill1974}, highlighting the main points of this theory. In fact, the analytical study of Gill was very thorough and complete with virtually all important theoretical questions arising in the linear theory answered.
Recent resurgence of interest to MI was motivated by important plasma and GFD applications
~\citep{Diamond2005,Manin1994,SmolyakovShev2000,Onishchenko2004,Onishchenko2008,Champeaux2001,Connaughton2010,Gallagher2012}.
A nonlinear analytical theory of MI for CHM model was developed by~\cite{Manin1994} using a scale separation technique.
Linear MI theory for EHM was first considered~\citep{Champeaux2001} using a scale separation technique similar to the
one of ~\cite{Manin1994}.
Recently, both linear and nonlinear MI theory within the CHM model was thoroughly revisited and a detailed numerical
study of the nonlinear MI development was carried out by~\cite{Connaughton2010}.
In a follow-up paper by~\cite{Gallagher2012} a similar study was performed for the EHM model.
Below we will overview and summarise these studies.

\subsection{Four-mode truncation}

Following the traditional setup of the MI problem, a meridional primary wave, $\bt{p}$, a small zonal perturbation or modulation, $\bt{q}$ and the two coupled sidebands $\bt{p}_{\pm}=\bt{p}\pm\bt{q}$ are the four excited modes used for analysis.  It has been shown however, that for small nonlinearity, only three modes are required to describe the MI growth but using four modes is more robust for all levels of nonlinearity~\citep{Connaughton2010}.  

Thus, we will truncate the system (\ref{eq-CHMk})
by allowing the wave vectors  $\bt{k}$, $\bt{k}_1$ and $\bt{k}_2$
take only values $\bt{p}$, , $\bt{q}$, $\bt{p}_{\pm}=\bt{p}\pm\bt{q}$ and their negatives (taking into account that $\psi_{-\bt{k}} = \overline \psi_{\bt{k}}$).
It is convenient to introduce interaction representation variables $\Psi_{\bt{k}}(t) = \psi_{\bt{k}}(t) {\rm e}^{i \omega_{\bt{k}}\, t}$.  Then in place of Eq.~(\ref{eq-CHMk}) we have the following system,
\begin{eqnarray}
\nonumber \partial_t \Psi_{\bt{p}}&=& T(\bt{p},\bt{q},\bt{p}_-)\, \Psi_{\bt{q}} \Psi_{\bt{p_-}} {\rm e}^{i\, \Delta_-\, t} \\
\nonumber & & + T(\bt{p},-\bt{q},\bt{p}_+)\, \overline{\Psi}_{\bt{q}} \Psi_{\bt{p_-}} {\rm e}^{i\, \Delta_+\, t}\\
\nonumber \partial_t \Psi_{\bt{q}} &=& T(\bt{q},\bt{p},-\bt{p}_-) \Psi_{\bt{p}} \overline{\Psi}_{\bt{p_-}} {\rm e}^{-i\, \Delta_-\, t}\\
\nonumber  && + T(\bt{q},-\bt{p},\bt{p}_+) \overline{\Psi}_{\bt{p}} \Psi_{\bt{p_+}} {\rm e}^{i\, \Delta_+\, t}\\
\label{4MT}\partial_t \Psi_{\bt{p}_-}&=& T(\bt{p}_-,\bt{p},-\bt{q}) \Psi_{\bt{p}}\, \overline{\Psi}_{\bt{q}} {\rm e}^{-i\, \Delta_-\, t}\\
\nonumber \partial_t \Psi_{\bt{p}_+}&=& T(\bt{p}_+,\bt{p},\bt{q}) \Psi_{\bt{p}}\, \Psi_{\bt{q}} {\rm e}^{-i\, \Delta_+\, t}\,.
\end{eqnarray}
where $\Delta_{\pm} = \omega_{\bt{p}} \pm \omega_{\bt{q}} -\omega_{\bt{p_{\pm}}}$.  
We will call system (\ref{4MT}) the four mode truncation (4MT) of the CHM/EHM models.

Strictly speaking, the chosen four modes $(\Psi_{\bt{p}},\Psi_{\bt{q}},\Psi_{\bt{p}_+},\Psi_{\bt{p}_-})$ are coupled to further modes and do not form a closed system.  Indeed, the linear problem only strictly closes with the inclusion of all the satellites $\pm \bt{q} + m \bt{p}$ where $m$ is an integer~\citep{Gill1974}.  However, in considering the linear instability it is traditional to truncate the system to the four modes only with a justification that the higher--order satellites are less excited
in the linear eigenvectors, which turns out to be a very good approximation for weak
primary waves and quite reasonable for strong ones (Gill 1974).

\subsection{Linear dynamics of modulations}

Let us linearise Eq.~(\ref{eq-CHMk}) about the pure primary wave solution.  Introducing the vector notation
$\bt{\Psi} = (\Psi_\bt{p},\Psi_\bt{q},\Psi_{\bt{p}_+},\Psi_{\bt{p}_-})$, a monochromatic
primary wave is given by $\bt{\Psi}_0 = (\Psi_0,0,0,0)$ where $\Psi_0$ is a complex constant representing the initial amplitude of the primary wave i.e. $\Psi_0=\Psi_{\bt{p}}|_{t=0}$ and is an exact solution of Eq.~(\ref{eq-CHMk}).  The idea is to determine how stable this solution is to small perturbations, comprised of modes $\bt{q}$, $\bt{p}_+$ and $\bt{p}_-$, by taking $\bt{\Psi}=\bt{\Psi}_0 + \epsilon \bt{\Psi}_1$ where $\bt{\Psi}_1=(0,\widetilde{\psi}_{\bt{q}},\widetilde{\psi}_\bt{p_+},\widetilde{\psi}_\bt{p_-})$.  Linearising Eqs~(\ref{4MT}) for the 4MT model at first order in $\epsilon$ gives
\begin{eqnarray}
\nonumber \partial_t \widetilde{\psi}_\bt{q} &=& T(\bt{q},\bt{p},-\bt{p}_-)\, \Psi_0\,\overline{\widetilde{\psi}}_\bt{p_-} {\rm e}^{-i\, \Delta_-\, t}\\
\nonumber&&+ T(\bt{q},-\bt{p},\bt{p}_+)\, \overline{\Psi}_0\,\widetilde{\psi}_\bt{p_+} {\rm e}^{i\, \Delta_+\, t} \\
\nonumber \partial_t \overline{\widetilde{\psi}}_{\bt{p}_+} &=& T(\bt{p}_+,\bt{p},\bt{q})\, \Psi_0\, \widetilde{\psi}_\bt{q}\, {\rm e}^{-i\, \Delta_+\, t}\\
\label{eq-linear4MT}  \partial_t \overline{\widetilde{\psi}}_{\bt{p}_-} &=& T(\bt{p}_-,\bt{p},-\bt{q})\, \overline{\Psi}_0\, \widetilde{\psi}_\bt{q}\, {\rm e}^{i\, \Delta_-\, t}.
\end{eqnarray}

Now solutions are sought of the form:
\begin{eqnarray}
\nonumber \widetilde{\psi}_\bt{q}(t) &=& A_\bt{q} {\rm e}^{-i\, \Omega_\bt{q}\,t}\,,\;\;\;\;
\widetilde{\psi}_{\bt{p}_+}(t) = A_{\bt{p}_+} {\rm e}^{-i\, \Omega_{\bt{p}_+}\,t}\\
& &\widetilde{\psi}_{\bt{p}_-}(t) =  A_{\bt{p}_-} {\rm e}^{-i\, \Omega_{\bt{p}_-}\,t},
\label{freq-match}
\end{eqnarray}
which requires for consistency that 

\begin{equation}
\Omega_{\bt{p}_+} = \Omega_\bt{q}+\Delta_+ \quad \hbox{ and} \quad \overline{\Omega}_{\bt{p}_-} = -\Omega_\bt{q}+\Delta_-.
\label{freq-match1}
\end{equation}
Writing Eqs~(\ref{eq-linear4MT}) in matrix format gives,
\begin{widetext}
\begin{equation}  
\label{matrixlinear4MTpm}
\begin{pmatrix} 
i\Omega_\bt{q} & T(\bt{q},-\bt{p},\bt{p}_+) \overline{\Psi}_0 & T(\bt{q},\bt{p},-\bt{p}_-) \Psi_0\\ 
T(\bt{p}_+,\bt{p},\bt{q}) \Psi_0 & i(\Omega_\bt{q}+\Delta_+)   & 0\\
T(\bt{p}_-,\bt{p},-\bt{q}) \overline{\Psi}_0 & 0 & -i(-\Omega_\bt{q}+\Delta_+)
\end{pmatrix}
\begin{pmatrix} 
A_\bt{q} \\ 
A_{\bt{p}_+} \\
\overline{A}_{\bt{p}_-} 
\end{pmatrix} = 0,
\end{equation}
and setting the determinant of the $3 \times 3$ matrix to zero, gives a cubic expression for the dispersion relation of the modulation frequency $\Omega_{\bt{q}}$,
\begin{eqnarray}
\nonumber\Omega_\bt{q}(\Omega_\bt{q}+\Delta_+)( -\Omega_\bt{q}+\Delta_-) + T(\bt{q},-\bt{p},\bt{p}_+)\, T(\bt{p}_+,\bt{p},\bt{q})\, \left|\Psi_0\right|^2 ( -\Omega_\bt{q}+\Delta_-)\\
\label{eq-MIDispersion}  - T(\bt{q},\bt{p},-\bt{p}_-)\, T(\bt{p}_-,\bt{p},-\bt{q})\, \left|\Psi_0\right|^2 ( \Omega_\bt{q}+\Delta_+)  = 0, \quad
\end{eqnarray}
\end{widetext}
which can be solved numerically for $\Omega_{\bt q}$, which is then used to find the corresponding eigenvectors
\begin{equation}
\label{4MTEigenvector}
\begin{pmatrix}
A_\bt{q}\\
A_{\bt{p}_+}\\
A_{\bt{p}_-}
\end{pmatrix} = 
\begin{pmatrix}
1\\
\frac{T(\bt{p}_+,\bt{p},\bt{q})\, \Psi_0}{-i\,(\Omega_\bt{q} + \Delta_+)}\\
\frac{T(\bt{p}_-,\bt{p},-\bt{q})\, \Psi_0}{i\,(\overline{\Omega}_\bt{q} - \Delta_-)}\,.
\end{pmatrix}.
\end{equation}
Instability occurs when a root $\Omega_{\bt q}$ has a positive  imaginary part, which is the growth rate of the instability, $\gamma = \Im (\Omega_{\bt q}) >0$. Then $\tau=\frac{1}{\gamma}$ is the characteristic growth time.  

\subsection{Nonlinearity strength parameter}

Let us introduce a dimensionless amplitude which measures the {\em initial} strength of the primary wave \citep{Gill1974,Connaughton2010}:
 \begin{equation}
M = \frac{\Psi_0 p^3}{\beta }\,,
\label{eq-M}
\end{equation}
where $\Psi_0=\Psi_{\bf{p}}|_{t=0}$.  
Parameter $M$ is a { formal} measure of nonlinearity in the MI evolution ie. the ratio nonlinear to linear terms.

\subsection{Cases with purely meridional carrier wave}

\begin{figure*}
\begin{center}
\includegraphics[width=\textwidth]{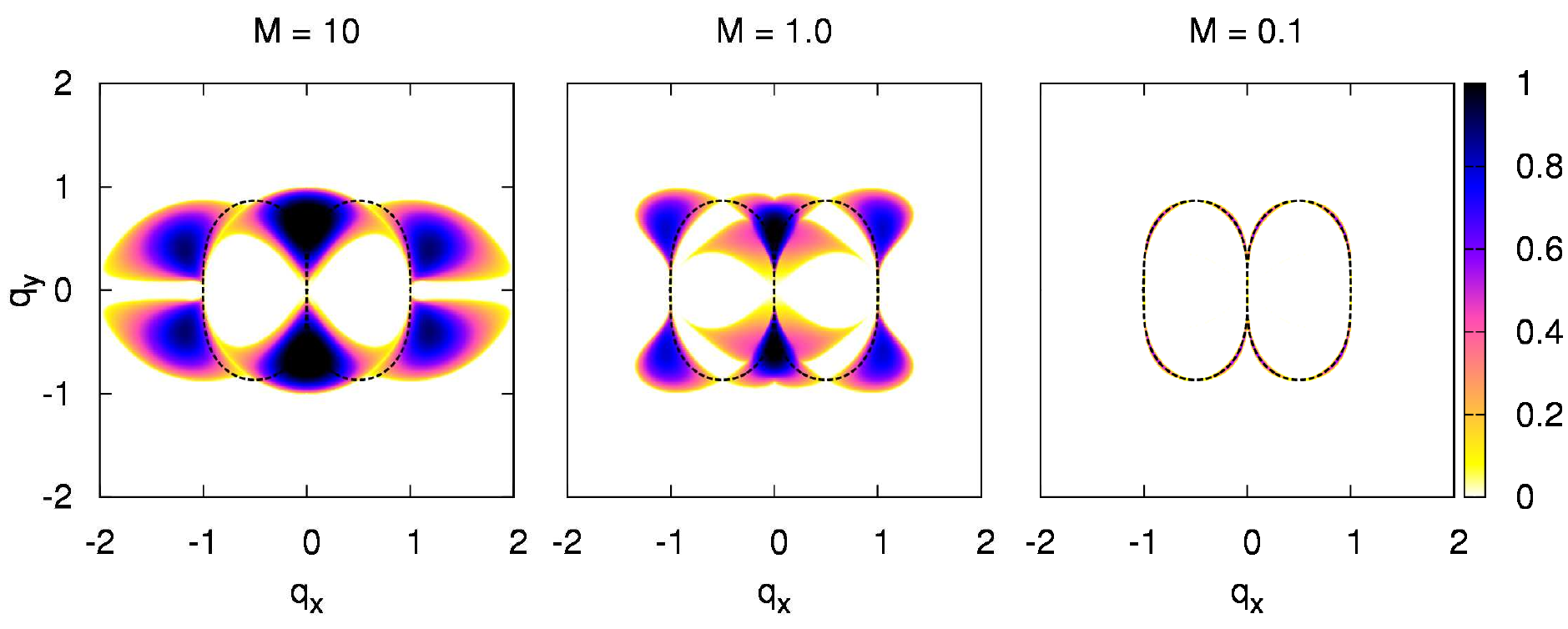}
\end{center}
\caption{\label{qMaps_F_0}  Growth rate of the modulational instability as a function of ${\bt q}$ for a fixed meridional primary wavevector, ${\bt p}=(1,0)$ and $\rho=\infty$ for various values of the nonlinearity $M$.}
\end{figure*}

The structure of the instability depends strongly on the parameter $M$.
For illustration, let us consider the case where the carrier wave is purely meridional, ${\bf p} = (p_x, 0)$. As seen in figure~\ref{qMaps_F_0}, for $M \to 0$ the instability concentrates narrowly around two oval-shaped curves. Each of these ovals appears to be the resonant curve on which the carrier wave, the modulation and one of the satellites are in exact three-wave resonance: 
\begin{eqnarray}
 {\bt p} = {\bt p}_- + {\bt q}, &\quad
\label{resonanceConditions}\omega({\bt p}) = \omega({\bt p}_-) +\omega({\bt q}) \\
\hbox{or} \quad 
 {\bt p} = {\bt p}_+ - {\bt q}, &\quad
\label{resonanceConditions1}\omega({\bt p}) = \omega({\bt p}_+) - \omega({\bt q}).
\end{eqnarray}

In the strong interaction limit, $M \to \infty$, the $\beta$-term in the CHM equation is unimportant and the MI problem is reduced to the well-known instability of the Kolmogorov (sine-shaped) plane-parallel shear flows described by the Euler equation for the incompressible fluid ~\citep{Arnold1960,Meshalkin1962}. The maximum instability is obtained when the primary and secondary waves are perpendicular~\cite{Gill1974} and the traditional approach is to select a meridional flow such that  $\bt{p}$ is along the $x$-axis and the perturbation $\bt{q}$ along the $y$-axis.  {That is to say, if the carrier wave is meridional, MI leads to the growth of zonal flows. Figure~\ref{qMaps_F_0} plots the growth rate of the perturbation having wave-vector $\mathbf{q}$ as a function of $\mathbf{q}$ in the case where the carrier wave-vector, $\mathbf{p}$, is meridional. Several values of $M$ are shown.}  For large and moderate values of $M$, the most unstable perturbation is clearly zonal. In the weak nonlinearity 
limit, however,  the maximally unstable perturbation bifurcates off the zonal axis at a critical value of $M_c \approx 0.53474$, tending to a point on the resonant manifold located at an angle of $5\pi/6$ with the $x$-axis. { Such modes are still predominantly zonal but are slightly inclined with respect to the $x$-axis. This is one possible mechanism for the weak inclined jets observed in simulations of the Earth's oceans in \cite{Maximenko2008}.}

\subsection{Cases with purely zonal modulation}


By fixing the modulation ${\bf q}$ to be purely zonal, one can study the instability for different { carrier wave-vectors, ${\bf p}$}. This results in the most familiar statement in the MI papers in plasmas that the primary wave will be unstable if it lies within the cone ~\citep{Manin1994,SmolyakovShev2000,Onishchenko2004}
$1/\rho^2 + p^2_x -3p^2_y > 0$.
When $\rho=\infty$, this cone reduces to $p_y < \frac{1}{\sqrt{3}} p_x$.  
However, this is true only in the double limit $M\gg1$ and $p\gg q$.

Removing the scale separation condition $p\gg q$ but still keeping $M\gg1$
we have a more general instability region~\citep{Gill1974}:
$\cos^2 \phi < ({1+ {q^2}/{p^2}})/{4},
$
where $\phi$ is the angle between $\bf{p}$ and $\bf{q}$.

\begin{figure*}
\begin{center}
\includegraphics[width=\textwidth]{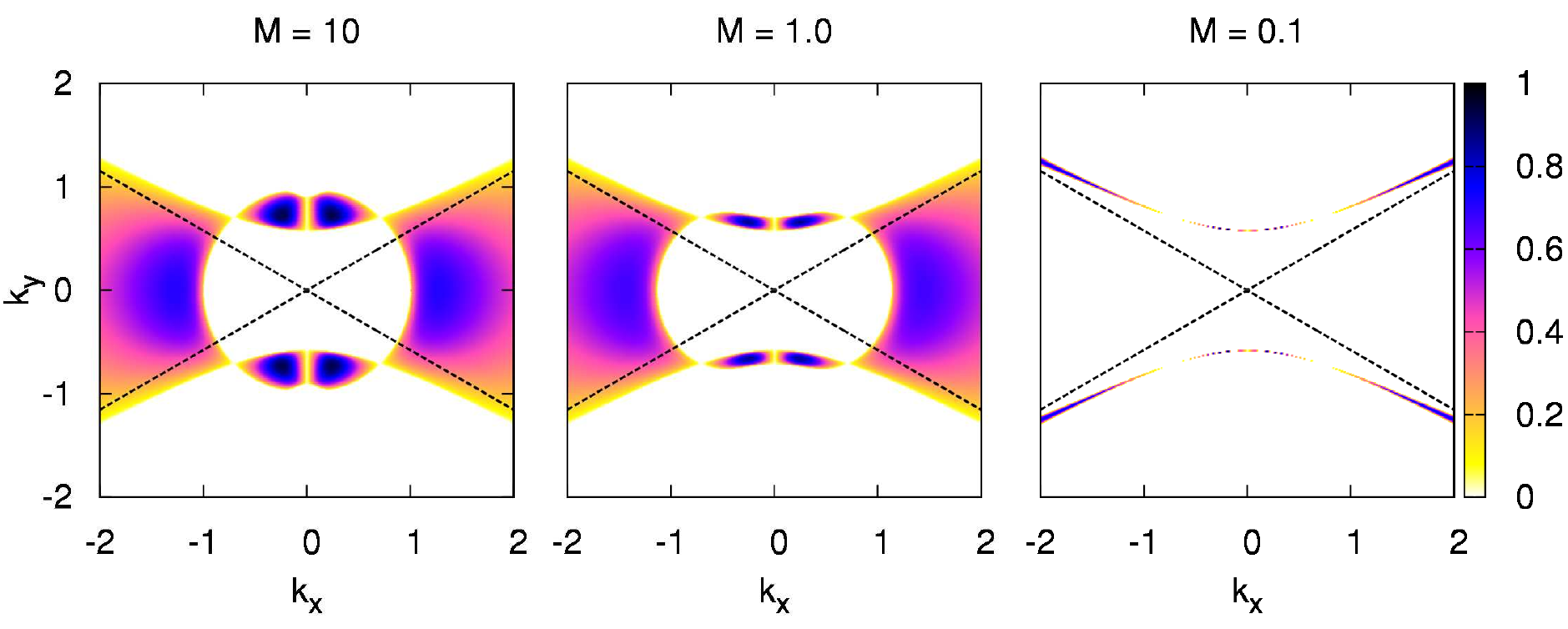}
\end{center}
\caption{\label{kMaps_F_0}
Growth rate of the modulational instability, given by  Eq.~\ref{eq-MIDispersion}
as a function of $\bf{p}$ for a fixed zonal modulation wavevector, $\bf{q}=(0,1)$ and $\rho=\infty$ for various levels of $M$. The dashed line is the cone defined by $p_y < \frac{1}{\sqrt{3}} p_x$}
\end{figure*}

Figure~\ref{kMaps_F_0} shows the instability growth rate 
 as a function of ${\bf p}$ for a fixed zonal modulation ${\bf q}=(0,1)$ and $F=0$ (ie. $\rho = \infty$). 
  As $M \to 0$, a region of stable wavenumbers inside the cone $p_y < \frac{1}{\sqrt{3}} p_x$ becomes larger such that unstable wavevectors require a larger $p_x$.  In the opposite limit of large $M$, an instability exists for some wavenumbers outside the cone which are very close to the zonal direction  
such that the maximum growth 
occurs for the primary wave orientations closer to zonal
 than to the the meridional direction, see Fig.~\ref{kMaps_F_0}  for $M=10$.
 
On the other hand, the choice of the primary wave direction is often dictated not by
 the maximum growth rate of the modulational instability, but by the structure of the
 primary instability creating the Rossby and drift waves.
 
In the limit of weak nonlinearity $M\ll 1$, the dynamics are completely wave dominated, and the instability is again close to the resonant curves
(\ref{resonanceConditions}) and (\ref{resonanceConditions1}).
An interesting feature of instability for $M \ll 1$ is evident in figure~\ref{kMaps_F_0},
that for fixed zonal ${\bf q}$ the unstable region becomes narrow and collapses onto the sides of the cone i.e. onto the lines $p_y = \pm p_x /\sqrt{3}$~\citep{Connaughton2010}. 

\subsubsection{Finite $\rho$ effects  and EHM}

First of all we recall that the EHM case is identical to CHM when the Larmour radius 
$\rho$ is infinite. The two systems are also identical unless one of the modes (most likely $\bf q$) is purely zonal.
Thus, all results described above for the cases $\rho = \infty$ or/and
when the modulation was not purely zonal apply equally to the CHM and the EHM models.

In CHM, when $\rho$
is finite, there are two regimes~\citep{Connaughton2010}, depending on the value of $M$:
For $M>\sqrt{\frac{2}{27}}$  the finite radius reduces the growth rate of the instability but cannot suppress it, whereas for
 $M \le \sqrt{\frac{2}{27}}$  there is a range of intermediate radii which completely suppress the instability.

Modulational instability within the EHM model was considered by~\cite{Champeaux2001}. Their WKB-type method assumes scale separation between the carrier wave and the modulation and (implicitly) strong nonlinearity $M \gg 1$ (see discussion in~\cite{Connaughton2010}).
It was shown that for EHM the instability is  stronger for purely zonal $\bf q$. 
This conclusion is generally true also for finite values of $M$~\citep{Gallagher2012}.

\begin{figure}
\begin{center}
\includegraphics[width=\columnwidth]{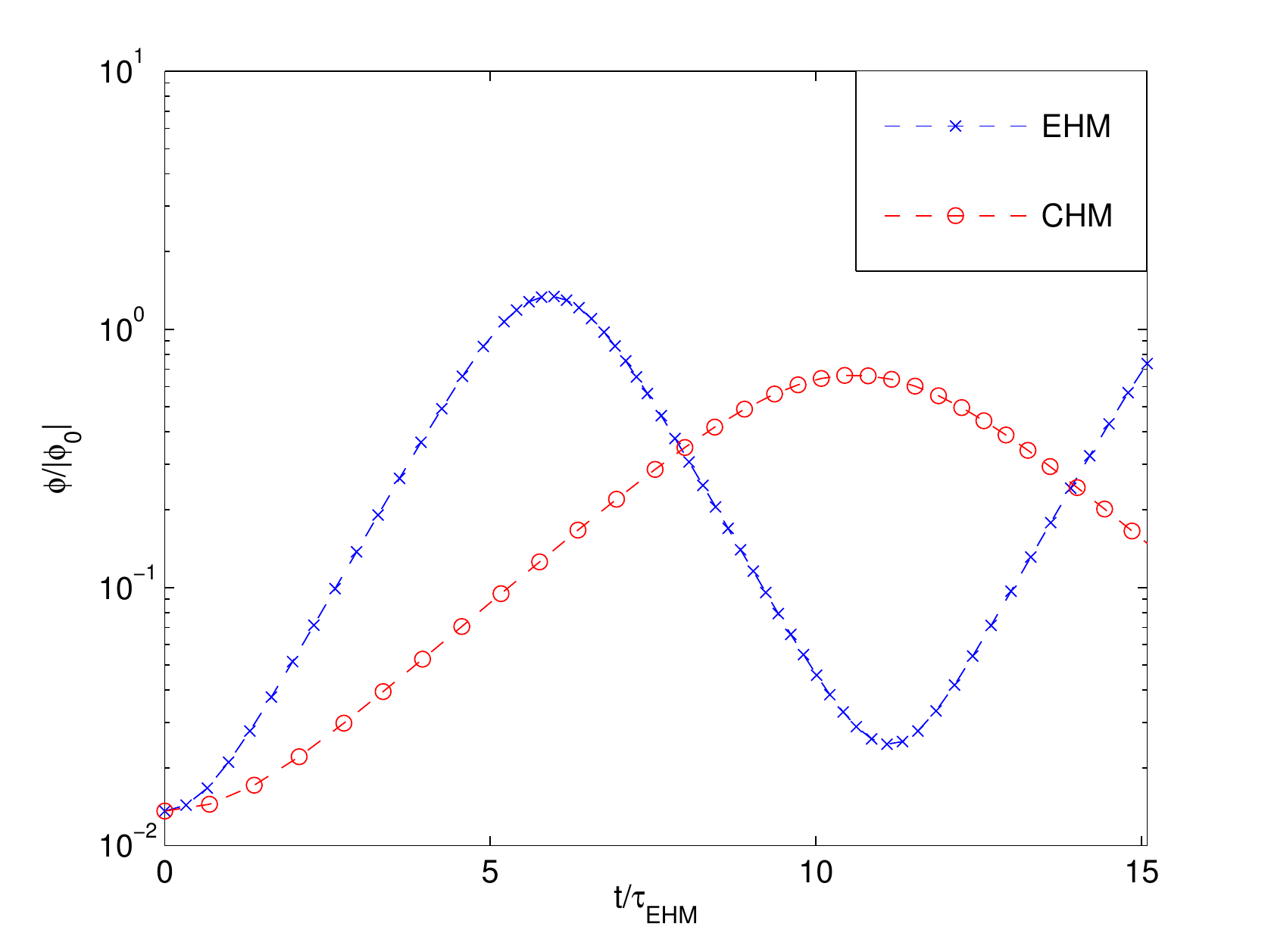}
\caption{The growth of the zonal mode for identical initial conditions in ODE simulations of the EHM and CHM
systems for the case $\bt{p}=(10,0)$, $\rho = 0.6$, $\psi_0=0.01$ and $\beta = 10 $. 
{ Times have been nondimensionalized with the linear growth rate of the EHM system.} }
\label{EHMgrowth}
\end{center}
\end{figure}

  Figure~\ref{EHMgrowth} shows the amplitude of the perturbation
mode $\bt{q}$ obtained by solving numerically the 4MT  system (\ref{4MT})  for the EHM and CHM equations.  Clearly, the effect of the extension to the CHM equation is that the growth rate associated with the EHM model is larger than that for the CHM model.

\section{Numerical results}



{In this section we compare the dynamics of the 4MT system, Eqs~(\ref{4MT}), with direct numerical simulations (DNS) of Eq.~(\ref{eq-Charney}).}  Although the 4MT was used as the departure point
for the linear stability analysis, it is a fully nonlinear set of equations
in its own right. In addition to checking the linear instability predictions
against DNS, the extent to which the nonlinear  dynamics
of the 4MT captures the behaviour of the full PDE will also be explored. In all cases, the initial condition is chosen to be along the unstable eigenvector of the 4MT.

\begin{figure*}
\includegraphics[width=\textwidth]{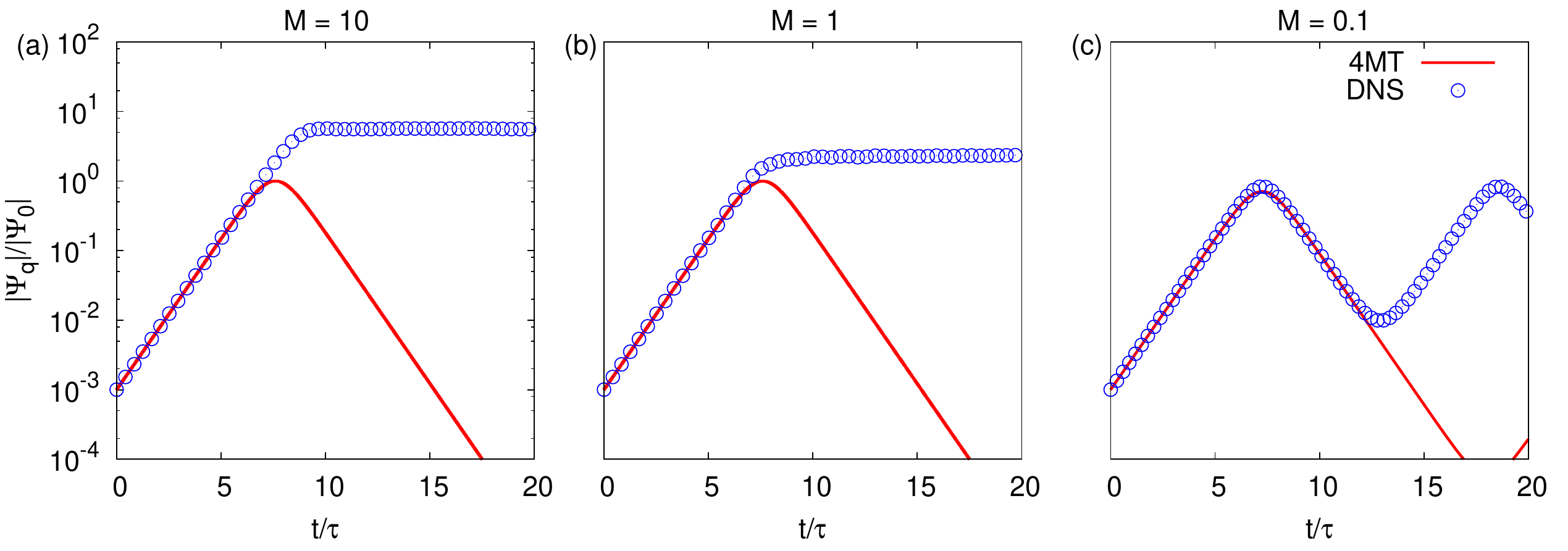}
\caption{Comparison of the growth of the zonal mode $\bf{q}$ obtained by DNS versus solving 4MT system.  In each case the primary wavenumber is ${\bf p} = (10,0) $ and the modulation wavenumber is ${\bf q} = (0,1) $.  The nonlinearity levels are (A) $M=10$, (B) $M=1.0$ and (C) $M=0.1$ and time is scaled by the corresponding $\tau$.}
\label{growth}
\end{figure*}

Immediately from figure~\ref{growth}, it can be seen that can see that the initial stage of evolution agrees very well with the predictions of the linear theory obtained from the 4MT.  Moreover, the 4MT works
rather well beyond the linear stage, particularly in the $M = 0.1$ case, where the initial
growth reverses in agreement with the (periodic) behaviour of the four-mode system.
For $M =1$, the  growth does not reverse, but rather experiences a saturation
at the level where the four-mode system reaches its maximum and reverses. The most
surprising behaviour is seen for M =10 where the linear exponential growth continues
well beyond the point of reversal of the four-wave system, even though the system is
clearly nonlinear at these times and follows a self-similar evolution (see below).

\begin{figure*} 
\includegraphics[width=\textwidth]{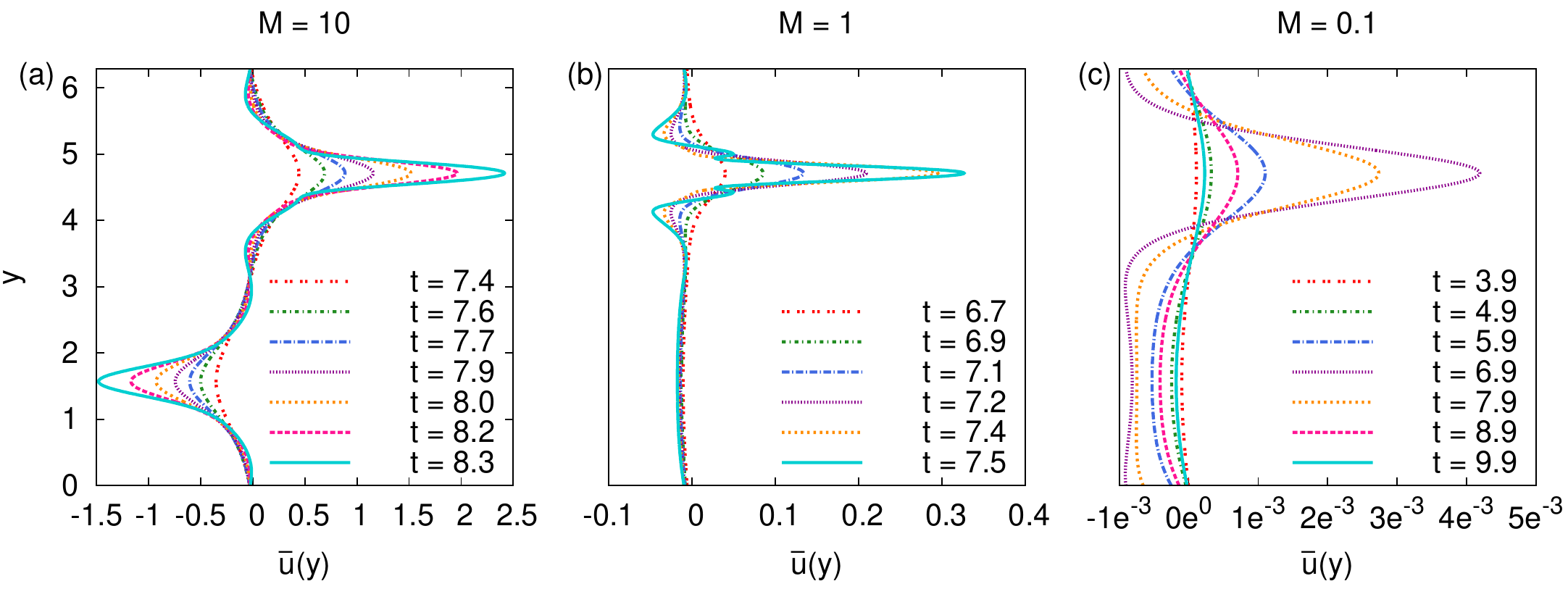}
\caption{Mean zonal velocity profiles for CHM ($s=0$); (a) $M=10$, (b) $M=1.0$ and (c) $M=0.1$.}
\label{zonalU}
\end{figure*}

A series of snapshots of the vorticity field for the nonlinearity levels $M=0.1$ 
and $M=10$ are shown in figures~\ref{snapshotsM0_1}
and~\ref{snapshotsM10} respectively.  The cases simulated are for ${\bf p} = (10,0) $ and ${\bf q} = (0,1) $.  The evolution of the mean zonal velocity $\overline u(y)$ averaged over $x$,
obtained from DNS is shown in figure~\ref{zonalU} for times close to the formation of the jet.
Below we will consider these figures in the context of three different stages of the nonlinear evolution.

\subsection{Initial nonlinear stage: sharpening of jets}

The nonlinear dynamics show zonal jets self-focusing and becoming very narrow with respect to the initial modulation wavelength.
This self-focusing was predicted theoretically~\citep{Manin1994} for large $M$ and $q \ll p$ where self-similar solutions were obtained describing a collapse of the jet width.  This feature cannot be described by the 4MT because such an harmonic jet shape involves strong contributions from higher harmonics ${\bf p} \pm n {\bf q}$. 

It was shown in \cite{Connaughton2010} that  the zonal velocity $\overline u$   in  a run with $M=10$ and $\rho = \infty$ can be fitted with a
self-similar shape $\overline u(y,t) = a(t) \, f(b(t) y)$, where $a(t) = u_0 \, e^{\gamma_{\bt q} t}$ and $b(t) = e^{1.85 t}$.  
The nonlinear growth at the self-similar stage continues with the same exponential law, ${\rm e}^{\gamma_{\bt q} t}$, as in the linear dynamics.  The self-similarity must stop when the scale separation property breaks down due to the jet narrowing, at which point
a roll-up into vortices occurs.
For smaller $M$, the extension of the growth rate beyond the linear stage is not observed and the amplitude of the zonal mode decreases after reaching a maximum in correspondence with the solution of the 4MT. The self-focusing is still observed but much reduced and the self-similar stage is not clearly seen.

\subsection{Intermediate nonlinear stage: vortex roll-up vs oscillation}

\begin{figure*}
\includegraphics[width=\textwidth]{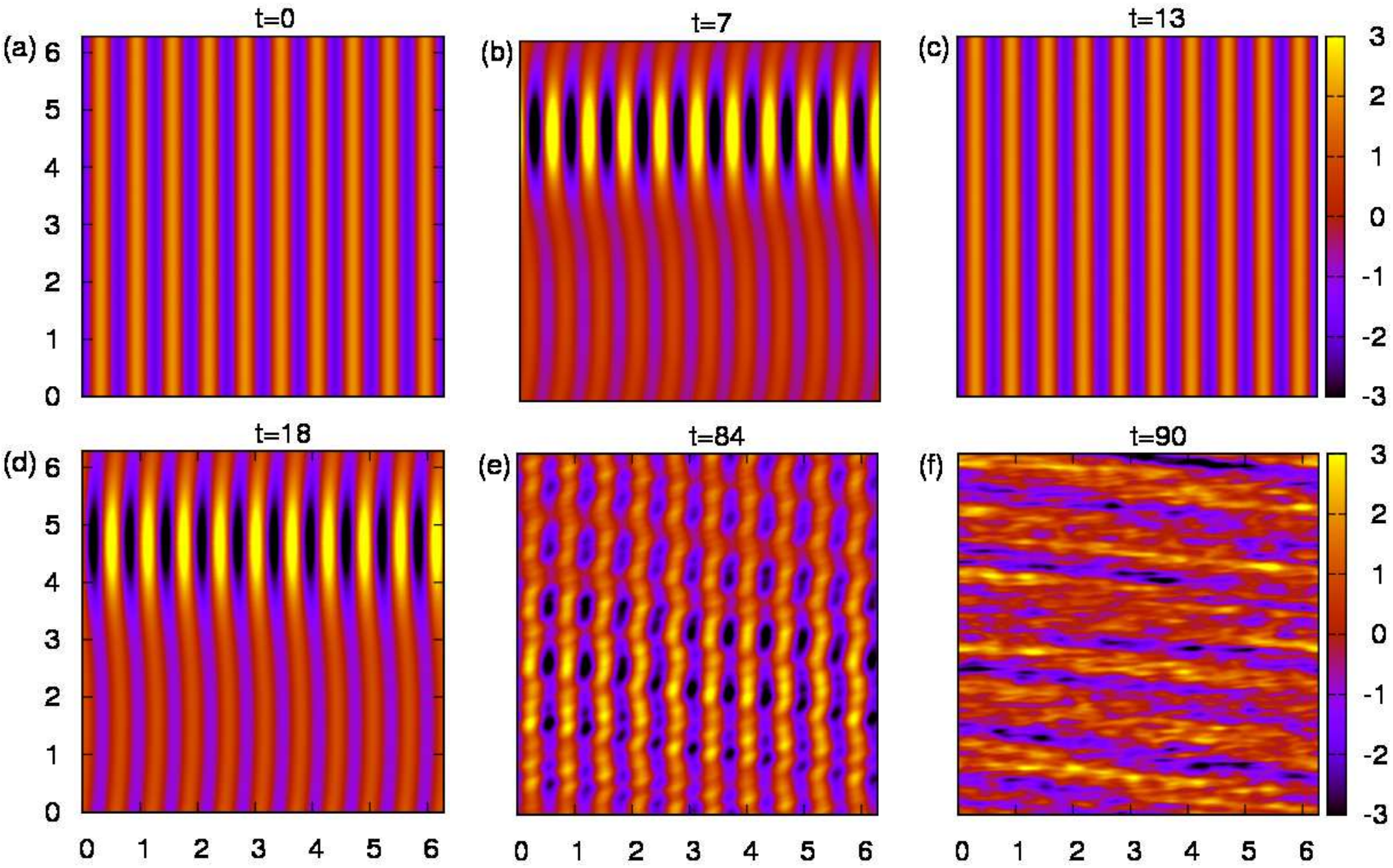}
\caption{Vorticity snapshots for a run with ${\bf p} = (10,0) $ and ${\bf q} = (0,1) $, $M=0.1$ and $\rho=\infty$. The growth stage is followed by growth reversal, saturation and transition to irregular off-zonal jet structures.
}
\label{snapshotsM0_1}
\end{figure*}
\begin{figure*}
\includegraphics[width=\textwidth]{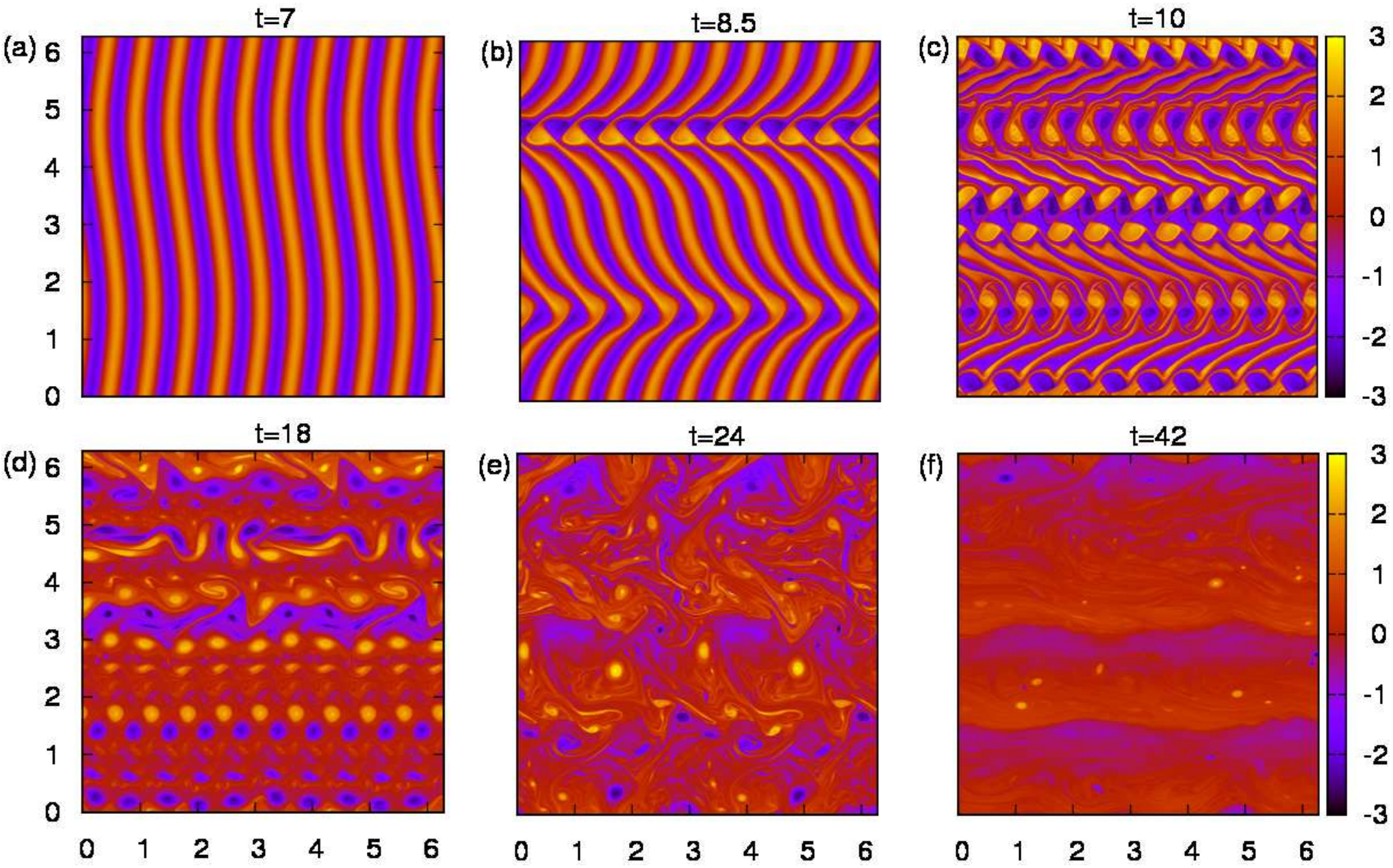}
\caption{Vorticity snapshots for a run with ${\bf p} = (10,0) $ and ${\bf q} = (0,1) $, $M=10$ and $\rho=\infty$. 
 The growth stage is followed by saturation to a vortex street state slowly deteriorating  to turbulence with a PV staircase structure.
  }
\label{snapshotsM10}
\end{figure*}

It is evident from figures~\ref{snapshotsM0_1}
and~\ref{snapshotsM10} that there are two qualitatively different types of behaviour, namely vortex roll-up followed by saturation or the oscillatory wave dynamics. 
For weak waves, $M < M_*$, the  oscillatory behaviour is seen where the system returns close to the initial state.  The 4MT model captures this effect very well.   This oscillation between a Rossby wave and a zonal flow has been reported from numerical studies of a barotropic fluid~\citep{Mahanti1981}, motivated by observations of such periodic behaviour in the troposphere of the southern hemisphere~\citep{Webster1975} and similar behaviour has also been noted for the EHM model~\citep{Manfredi2001} for drift turbulence.
For strong waves, $M > M_*$, the nonlinear evolution is vortex dominated and the vorticity of the initial primary wave rolls into vortices and organises itself into K\'arm\'an-like vortex streets.  This corresponds to the jet velocity saturating. At the moment of the vortex roll-up energy spills beyond the four modes and the 4MT description fails.

 It is therefore natural to seek the critical level of nonlinearity $M_* $ which distinguishes between the two types of behaviour.  If the maximum jet strength, as predicted by the 4MT, exceeds the value of the Rayleigh-Kuo necessary instability condition $\partial_{yy} \overline u(y) - \beta >0$~\citep{Rayleigh1913,Kuo1949,McWilliams2006,Connaughton2010}, then the vortex roll-up occurs and the jet strength saturates for a long time. 
 At this point, the behaviour of the system starts to depart from that of the 4MT.  If however, the maximum jet strength remains below the Rayleigh-Kuo threshold, then the system's growth reverses and follows the 4MT dynamics for a longer time.

This simple picture permits a qualitative physical estimate for the saturated velocity of the jet  $u_{max}  \approx 3 \frac{M \beta } {p^2}$ and respectively $M_* \sim \frac{1}{3}$, which agrees with the DNS result for the oscillatory-saturated transition at   $M_* \approx 0.25-0.35$~\citep{Connaughton2010}.

\subsection{Late nonlinear stage: turbulence}

In all cases the orderly nonlinear motion of the intermediate stages
is followed by disordered turbulent states at the late evolution stages.
However, the properties of such turbulence are different for
the cases {$M < M_*$} and {$M > M_*$}, as clearly seen in the last
frames of the figures~\ref{snapshotsM0_1}
and~\ref{snapshotsM10}.


For weak waves, $M < M_*$, the  quasi-oscillatory behaviour where the system periodically returns close to the initial state is not sustained and a transition to an anisotropic wave turbulence state occurs.
The dominant jet structures observed in such a turbulent state in figure~\ref{snapshotsM0_1}(f) are
off-zonal. This effect may be connected to the off-zonal ``striations'' reported
for ocean observations~\citep{Maximenko2008} although these ocean striations only become evident in the averaged data since they are sufficiently weak.

For strong waves, $M > M_*$,
at the final stages the vortex streets break up due
to a vortex pairing instability which is followed by a transition to vortex turbulence.  Such turbulence is anisotropic with a pronounced zonal jet component and a well-formed potential vorticity (PV) staircase is evident in figure~\ref{snapshotsM10}(f)~\citep{Dritschel2008}.

\subsection{Role of the deformation radius}

\begin{figure*}
\includegraphics[width=\textwidth]{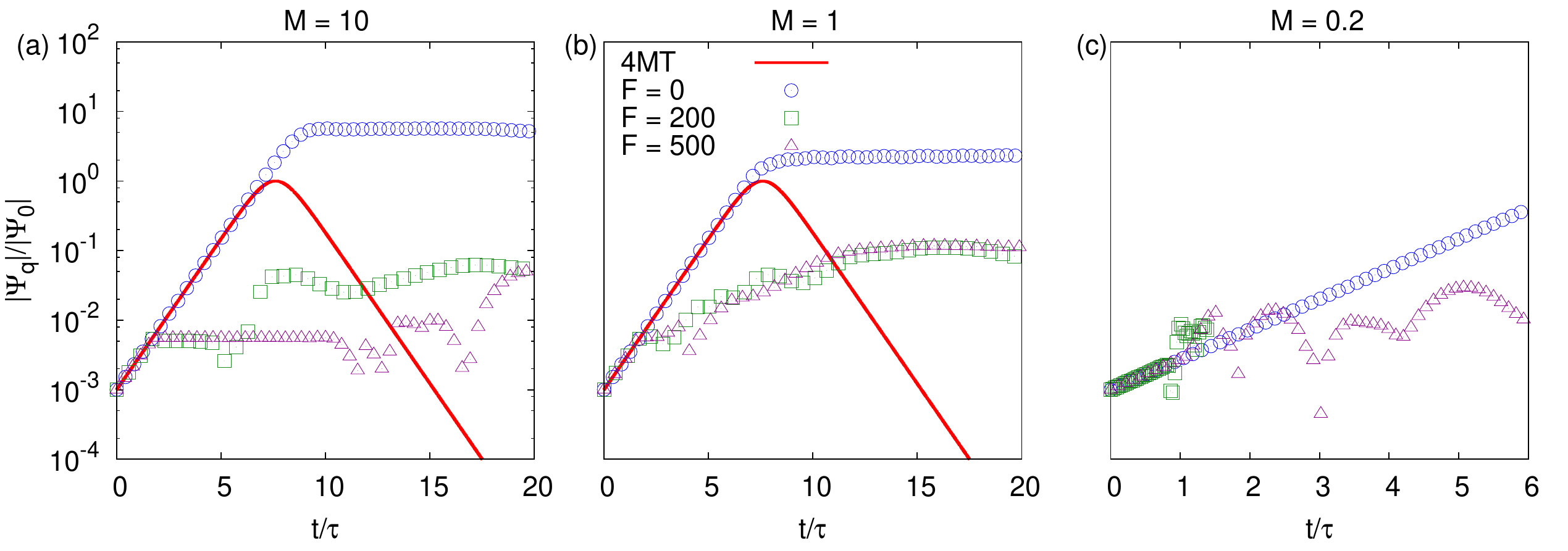}
\caption{Comparison of the growth of the zonal mode $\bf{q}$ obtained by DNS versus solving 4MT system for various $F=\frac{1}{\rho^2}$.  In each case the primary wavenumber is ${\bf p} = (10,0) $ and the modulation wavenumber is ${\bf q} = (0,1) $.  The nonlinearity levels are (a) $M=10$, (b) $M=1.0$ and (c) $M=0.2$.}
\label{growthF}
\end{figure*}

{ Until now we have focused mostly on the case of infinite deformation radius, $\rho$, corresponding to $F=0$. We now discuss the effect of finite $F$. In section \ref{sec-MI} we used the wavelength of the carrier wave, $2\pi/\left|\mathbf{p}\right|$, as our unit of length when nondimensionalizing Eq.~(\ref{eq-Charney}). Therefore in what follows, $\rho=1$ corresponds to a deformation radius comparable to the carrier wavelength.}  When $\rho$ is finite, some vortex streets are still evident.
In fact, vortex roll-up occurs quicker at the stage where the 4MT model still predicts a linear growth i.e. well before the 4MT prediction for the zonal mode reaches its maximum.
This is evident in figure~\ref{growthF} where the predicted linear growth rate is now observed for only one to two timescales.  

\subsection{EHM}

Numerical results also show that two qualitatively distinct regimes, saturation versus oscillation, are also observed by varying $\rho$ in the EHM model.  
\begin{figure}
\includegraphics[width=\columnwidth]{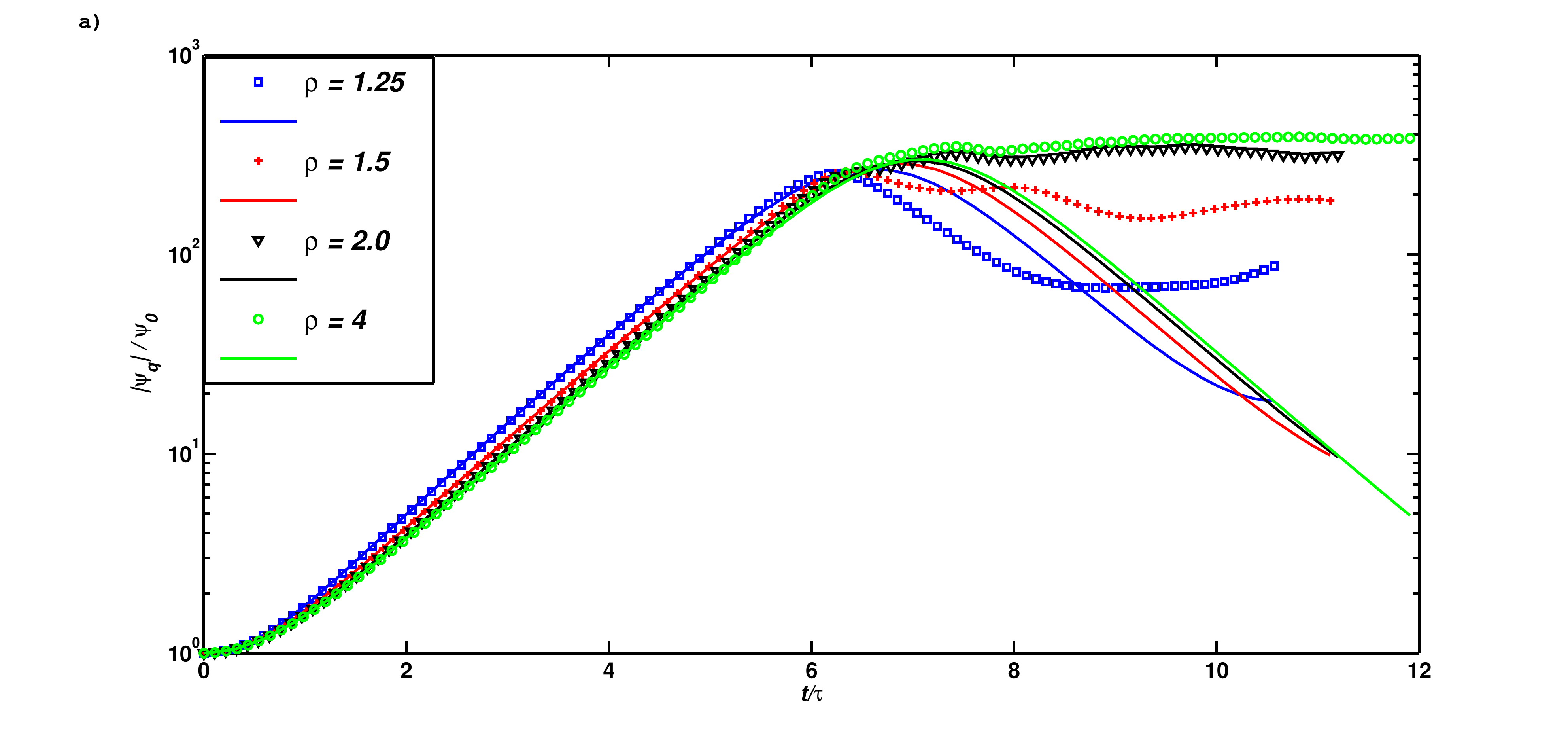}
\includegraphics[width=\columnwidth]{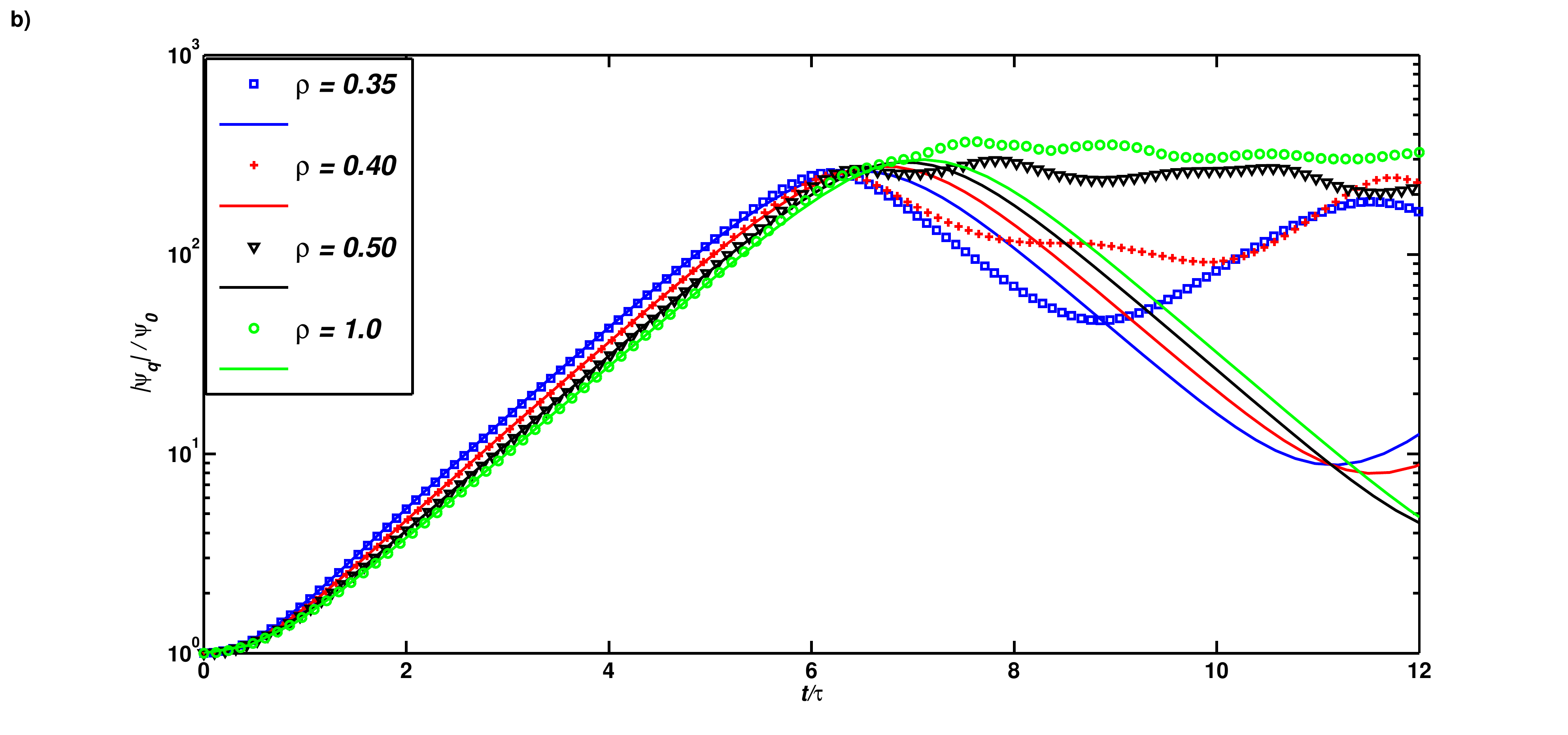}
\caption{The amplitude of the zonal mode for various $\rho$ with (a) $\psi_0 = 0.001$ ($M=0.1$) and (b) $\psi_0 = 0.01$ ($M=1$), other parameters were identical, $\beta =10$, $\bt{p}=(10,0)$, $\bt{q}=(0,1)$. Each case has been scaled by its own linear growth time. ODE predictions are shown with solid lines, full simulations with markers. }
\label{EHM_Mcrit}
\end{figure}
Figure \ref{EHM_Mcrit}(a) shows the amplitude of the zonal mode for several
different values of $\rho$ at the same value of the nonlinearity parameter $M=0.1$.  The runs all have a linear phase in which they grow at the rate predicted by Eq.~\eqref{eq-MIDispersion} yet increasing $\rho$ gradually changes the behaviour of the system so that the zonal mode
saturates rather than oscillates,  similar to CHM. This change in behaviour occurs between $\rho = 1.5$ and $\rho =2$. Note that transition to the saturated regime occurs in EHM at smaller levels of the nonlinearity than in the CHM model. Indeed, in CHM the case $M=0.1$ exhibited an oscillitary behaviour even for $\rho =\infty$. Figure \ref{EHM_Mcrit}(b) shows the second case in which  the initial
pump wave amplitude was ten times larger, $M=1$. In this case the switch from saturating to oscillatory behaviour occurs between $\rho = 0.4$ and $\rho =0.5$. 


\section{Instability and  zonal jets in HW/EHW model}

The HW/EHW model 
\eqref{NonLocalHWChap-HW-N} has a great advantage over the CHM/EHM in that it contains a primary instability capable of spontaneously generating of waves.
On the other hand, this very property makes this system unsuitable for the standard
MI approach based on 4MT. Namely, if we start with a single monochromatic
pump wave, perturb it with a modulation and two side-band modes and linearise the system, we would not be able to  find find a solution to such a linear system in which the frequencies of the modulation and the side-bands would match like in Eqs.~(\ref{freq-match}) and (\ref{freq-match1}) (which would yield a linear algebraic system with time-independent coefficients).

However, the 4MT approach still appears to be useful for analysis, especially of the initial evolution stages. 
Thus, we will truncate the system~\eqref{NonLocalHWChap-HW-N}
by allowing the wave vectors  $\bt{k}$, $\bt{k}_1$ and $\bt{k}_2$
take only values $\bt{p}$, , $\bt{q}$, $\bt{p}_{\pm}=\bt{p}\pm\bt{q}$ and their negatives (taking into account that $\psi_{-\bt{k}} = \overline \psi_{\bt{k}}$).  Then in place of Eqs.~\eqref{NonLocalHWChap-HW-N} we have the following system,
\begin{widetext}
\begin{eqnarray}
\nonumber 
\partial_t \hat{\psi}_\bt{p} &=& 
- \frac{\alpha z}{p^2} (\hat{\psi}_\bt{p} - \hat{n}_\bt{p})  +
 T(\bt{p},\bt{q},\bt{p}_-)\, \hat{\psi}_{\bt{q}} \hat{\psi}_{\bt{p_-}} 
+  T(\bt{p},-\bt{q},\bt{p}_+)\, \overline{\hat{\psi}}_{\bt{q}} \hat{\psi}_{\bt{p_-}} \,,\\
\nonumber \partial_t \hat{\psi}_{\bt{q}} &=& 
- \frac{\alpha z}{q^2} (\hat{\psi}_\bt{q} - \hat{n}_\bt{q})  +
T(\bt{q},\bt{p},-\bt{p}_-) \hat{\psi}_{\bt{p}} \overline{\hat{\psi}}_{\bt{p_-}} 
+ T(\bt{q},-\bt{p},\bt{p}_+) \overline{\hat{\psi}}_{\bt{p}} \hat{\psi}_{\bt{p_+}} \,,\\
\nonumber
\partial_t \hat{\psi}_{\bt{p}_-}&=& 
- \frac{\alpha z}{p_-^2} (\hat{\psi}_\bt{p_-} - \hat{n}_\bt{p_-})  +
T(\bt{p}_-,\bt{p},-\bt{q}) \hat{\psi}_{\bt{p}}\, \overline{\hat{\psi}}_{\bt{q}} \,,\\
 \label{4MT-HW-psi}
 \partial_t \hat{\psi}_{\bt{p}_+}&=& 
- \frac{\alpha z}{p_+^2} (\hat{\psi}_\bt{p_+} - \hat{n}_\bt{p_+})  +
T(\bt{p}_+,\bt{p},\bt{q}) \hat{\psi}_{\bt{p}}\, \hat{\psi}_{\bt{q}} \,.
\end{eqnarray}

\begin{eqnarray}
\nonumber
\partial_t \hat{n}_\bt{p} = - i \kappa p_y \hat{\psi}_\bt{p} + z \alpha 
 (\hat{\psi}_\bt{p} - \hat{n}_\bt{p}) \!  -\!\!
 R(\bt{p}, \bt{q}) 
(\hat{n}_\bt{p_-}  \hat{\psi}_\bt{q} - \hat{n}_\bt{q}  \hat{\psi}_\bt{p_-} 
+
\hat{n}_\bt{p_+}  \overline{\hat{\psi}}_\bt{q} -
 \hat{\psi}_\bt{p_+}  \overline{\hat{n}}_\bt{q}) \,,
\\
\nonumber
\partial_t \hat{n}_\bt{q} = - i \kappa q_y \hat{\psi}_\bt{q} + z \alpha 
 (\hat{\psi}_\bt{q} - \hat{n}_\bt{q}) \!  -\!\!
 R(\bt{p}, \bt{q}) 
(\hat{n}_\bt{p}  \overline{\hat{\psi}}_\bt{p_-} - \hat{\psi}_\bt{p}  \overline{\hat{n}}_\bt{p_-}
-\overline{\hat{n}}_\bt{p}  {\hat{\psi}}_\bt{p_+} +
\overline{\hat{\psi}}_\bt{p}  {\hat{n}}_\bt{p_+} 
)\,,
\\
\nonumber
\partial_t \hat{n}_\bt{p_-} = - i \kappa p_{-,y} \hat{\psi}_\bt{p_-} + z \alpha 
 (\hat{\psi}_\bt{p_-} - \hat{n}_\bt{p_-}) +
 R(\bt{p}, \bt{q}) 
(\hat{n}_\bt{p}  \overline{\hat{\psi}}_\bt{q} - 
\hat{\psi}_\bt{p}  \overline{\hat{n}}_\bt{q})
\,, \hspace{2.3cm}
\\
\partial_t \hat{n}_\bt{p_+} = - i \kappa p_{+,y} \hat{\psi}_\bt{p_+} + z \alpha 
 (\hat{\psi}_\bt{p_+} - \hat{n}_\bt{p_+}) -
 R(\bt{p}, \bt{q}) 
(\hat{n}_\bt{p}  {\hat{\psi}}_\bt{q} - 
\hat{\psi}_\bt{p}  {\hat{n}}_\bt{q})
\,, \hspace{2.3cm}
\label{4MT-HW-N}
\end{eqnarray}
\end{widetext}

\subsection{Primary (drift-dissipative) instability in the HW/EHW system}

Linearising Eq.\eqref{NonLocalHWChap-HW-N} about a plane wave solution, $\hat \psi_{\bf k},  \hat n_{\bf k} \sim e^{-i \omega_{\bf k} t}$, we get
\begin{eqnarray}
\omega_{\bf k}=\frac{1}{2}[-ib\pm ib(1-4i\omega_{*}/b)^{1/2}],
\end{eqnarray}
where $b=\alpha z (1+(\rho k)^{-2})$, and 
$\omega_{*}=k_{y} \kappa/(1+\rho^2 k^{2})$.
The plus sign is an unstable mode $\gamma_{\bf k} = \Im \omega_{\bf k} >0$ with maximum growth rate at ${\bf k} \eqsim (1/\rho, 0)$.


\subsection{Secondary instability in the HW system - MI}

The growth rate of the primary instability $\gamma_{\bf k}$ has a rather wide distribution around its maximum. The developing MI is consequently different than in systems with very narrow $\gamma_{\bf k} $. With narrow $\gamma_{\bf k} $ one can consider there to be a dominant carrier wave that becomes
unstable. With wider $\gamma_{\bf k} $ it is more natural to assume that modes close to the instability
maximum have comparable strengths. In terms of the 4MT this means that the modes ${\bf p} $, ${\bf p}_+ $ and ${\bf p}_- $
have similar amplitudes and growth rates whereas the mode ${\bf q} $ is initially weak.
Since ${\bf p} $, ${\bf p}_+ $ and ${\bf p}_- $ must be close to the primary instability maximum, they must also be close to each other. Therefore, there arises a scale separation $q \ll k$. Furthermore, for nearly meridional 
${\bf p} $, ${\bf p}_+ $ and ${\bf p}_- $ with $\bf q$ close to the resonant curve, $\bf q$ is nearly zonal. 
We will see later that the fastest growing
$\bf q$ is the smallest zonal wave number allowed by the box.

Thus, let us therefore consider mode $\bf q$ in Eqs.~\eqref{4MT-HW-psi} and:
(i) take into account
that all zonal $\bf q$ are neutrally stable to the primary instability mechanism, 
 $\gamma_{\bf q} = 0$, ie. that the linear terms would vanish from the mode-$\bf q$ equation,
(ii)
substitute into the right-hand side
$\hat \psi_{\bf p} = \psi_{\bf p}(0) e^{-i\omega_{\bf p} t},\, \hat \psi_{{\bf p}_-} = \psi_{\bf p}(0) e^{-i\omega_{{\bf p}_-}  t},\, \hat \psi_{{\bf p}_+}  = \psi_{\bf p}(0) e^{-i\omega_{{\bf p}_+} t}$, 
(iii) take into account that the real parts of the frequencies of the triads ${\bf p} $, ${\bf p}_+ $, ${\bf q} $  and ${\bf p} $, ${\bf p}_- $, ${\bf q} $
are in near resonance and that for the imaginary part we have $\gamma_{\bf p} \approx  \gamma_{\bt{p_-}} \approx  \gamma_{\bt{p_+}} $.
This gives:
$$
\partial_t \hat{\psi}_{\bt{q}} =
\lambda \, e^{2 \gamma_{\bf p} t},
$$
where
\begin{equation}
\label{lambda}
\lambda= T(\bt{q},\bt{p},-\bt{p}_-) \hat{\psi}_{\bt{p}}(0)  \overline{\hat{\psi}}_{\bt{p_-}}(0)  
+ T(\bt{q},-\bt{p},\bt{p}_+) \overline{\hat{\psi}}_{\bt{p}}(0)  \hat{\psi}_{\bt{p_+}}(0)
\end{equation}
which has a solution:
\begin{equation}
\hat{\psi}_{\bt{q}} =\hat{\psi}_{\bt{q}} (0) +
\frac{\lambda}{2 \gamma_{\bf p} } \, e^{2 \gamma_{\bf p} t}.
\end{equation}
Thus the MI  growth rate for the zonal mode in the HW/EHW system is simply the double of the maximum growth rate
of the primary instability. Note that this growth occurs only after a delay needed for the second term in Eq.~\eqref{lambda} to overtake the first one (the initial condition).

\subsection{Numerical Results}\label{sec.EHW_numerical_results}

As with the CHM and EHM models, the 4MT for the EHW Eqs~\eqref{4MT-HW-psi}~and~\eqref{4MT-HW-N}, was compared to the direct numerical simulations (DNS) of the full system \eqref{NonLocalHWChap-HW-N} 
For comparison with the analytical predictions, we select optimal 4MT sets that are most prominent in the DNS data after some initial adjustment time.
The simulations are split into two classes: the one close to the CHM/EHM limit where $\alpha \gg 1$ and the genuine  HW regime where $\alpha \sim 1$.

Firstly a case with $\kappa =10$ and $\alpha =10$ was considered for which the EHW system is close to the EHM. Low amplitude random noise was generated at the grid scale and used as the initial condition for both density and potential. After initial evolution the modes where the primary instability is near maximum have developed and by a beating mechanism they have generated a zonal mode.  In the language of our 4MT, 
the mode with greatest growth is ${\bf p}$, 
and the satellite modes ${\bf p}_+$ and ${\bf p}_-$ grow at approximately the same rate
provided $q\ll p$.
From the Fourier space data generated by DNS we select a set of four dominant modes, namely ${\bf p}=(0,\frac{18 \pi}{L})$, ${\bf q}=(\frac{2\pi}{L},0)$ and ${\bf p}_\pm = {\bf p} \pm {\bf q}$ were chosen for comparison to the reduced model as they satisfied the  condition $q\ll p$ and had  relatively strong amplitudes.
The 4MT system was initialised using the values from the full DNS at $t=0$.

\begin{figure}
\begin{center}
\includegraphics[width=0.8\columnwidth]{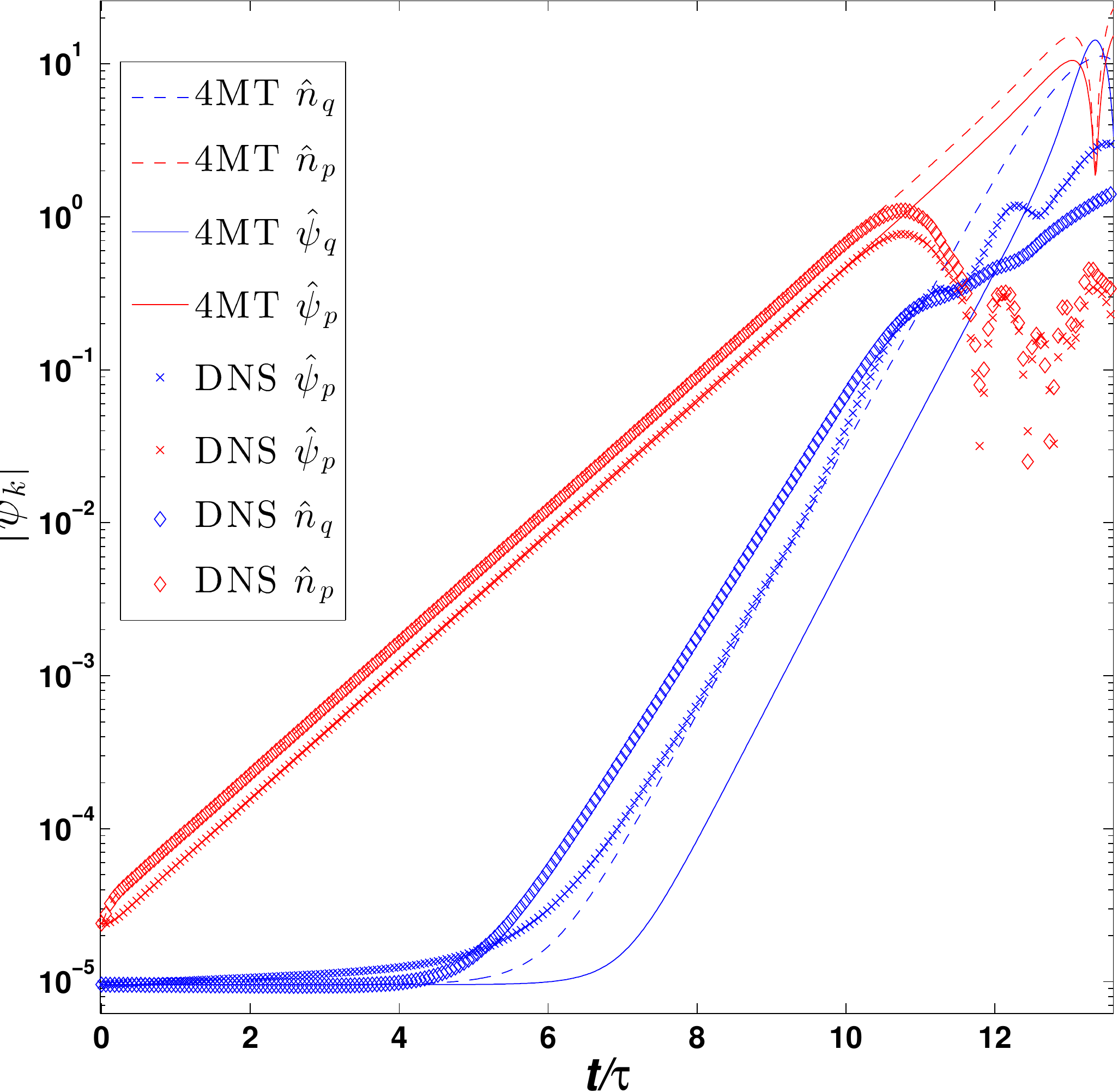}

\includegraphics[width=0.8\columnwidth]{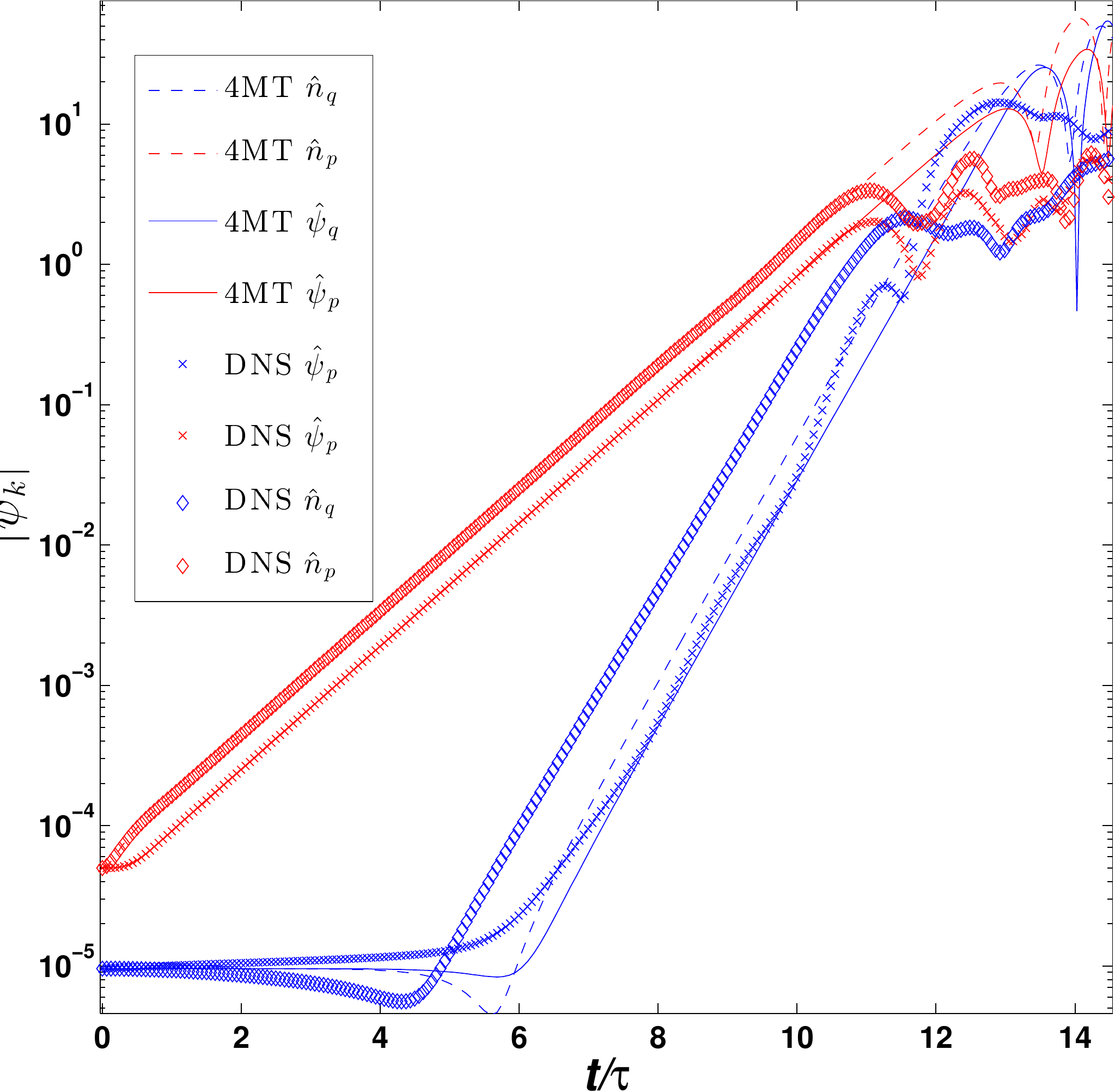}
\caption{A comparison between the 4MT predictions (lines) and the DNS  (markers) for (a) $\alpha=10$ case and (b) $\alpha=0.5$ case. 
\label{ODEvsPDE_A10}}
\end{center}
\end{figure}

The results of these simulations can be seen in figure \ref{ODEvsPDE_A10} (a) where amplitudes measured from the full DNS are compared with solutions of the 4MT equations. 
It can be seen that at very early times the density for mode ${\bf p}$ grows more rapidly than the potential until they align with the eigenvector of the primary instability, at which point they start growing synchronously exponentially with the growth rate predicted by the primary instability. The 4MT system captures this behaviour perfectly. In both DNS and in 4MT, the zonal modes, $\hat \psi_{\bf q}$ and $\hat n_{\bf q}$, show almost no growth until $t\sim 5\tau$ when they begin to grow exponentially at approximately double the linear growth rate of mode ${\bf p}$. The exponential  growth phase continues for both ${\bf p}$ and ${\bf q}$ until $t\sim 11\tau$, at which time the energies of these modes become comparable.
The 4MT agrees with DNS very well in all the qualitative and even quantitative features of the evolution, including the initial delay in the mode ${\bf q}$ growth  and the subsequent growth at the double rate up until the very advanced stage characterised by strong nonlinearity and turbulence. 
However, the 4MT system predicts slightly longer initial delay in the ${\bf q}$-mode
growth, which is most likely because during the initial evolution the selected modes ${\bf p}$, ${\bf p}_+$ and ${\bf p}_-$ are not so dominant with respect to the other modes driven by the primary instability and not encuded into the 4MT set; these modes provide extra pumping to the 
${\bf q}$-mode in the DNS.

The same procedure as above was carried out  with $\alpha=0.5$ and $\kappa=10$, which is a genuine EHW setup far from the EHM system. This range of parameters is known to give a reasonable qualitative agreement between simulated turbulence and that observed in fusion plasma~\citep{Hortonreview,Dudok1995}.  { A good choice of carrier wave and perturbation wave-vector to explore the correspondence with the 4MT are ${\bf p}=(0,-\frac{6 \pi}{L})$ and ${\bf q}=(\frac{2\pi}{L},0)$.} The 4MT system is compared to a full DNS in figure~\ref{ODEvsPDE_A10}(b). We see the same qualitative features as described for the case $\alpha=10$
and even better agreement between the 4MT and the DNS (the mismatch in the ${\bf q}$-mode delay appears to be less than before).

\section{Closing the loop: WT-ZF interaction}


MI is important for the initial evolution of turbulence, not only in conservative systems like CHM and EHM, but also in systems with external or spontaneous forcing, like forced CHM/EHM or HW/EHW respectively.
This mechanism becomes pronounced when the system is forced by a primary instability whose growth rate has a maximum at a meridional wavenumber (typically at $k \sim 1/\rho$), so that   initially a dominant quasi-monochromatic wave component appears.

After the dominant wave undergoes MI and a strong ZF component is generated, it is natural to ask how such a ZF will feed back
onto the instability scales and modify their dynamics.  
It is now generally accepted in both the fields of plasma physics and GFD that in the turbulence of Rossby and drift waves the dominant interaction is with a zonal flow rather than a neighbouring-scale interaction~\citep{Nazarenko051990,Nazarenko081990,Diamond2005,Shats2005,Onishchenko2008,Nazarenko2011}, such that studies of drift wave turbulence now imply the study of the drift wave turbulence (WT) - ZF feedback mechanism \citep{Nazarenko051990,Nazarenko081990,Nazarenko1991,Biglari1990,Diamond2005}.  
 
A summary of this mechanism is as follows. Rossby waves and drift waves are produced by a primary instability which is predominantly meridional,  such asthe ion-temperature-gradient (ITG) instability in plasmas or the baroclinic instability in GFD. ZFs are generated via a secondary modulational instability of these waves and they grow by a direct interaction with the small scale waves.  The growing ZF extracts energy from the wave turbulence (WT) and thereby eventually suppresses it at the small scales.


The WT-ZF feedback mechanism was first described theoretically in~\cite{Nazarenko051990,Nazarenko081990,Nazarenko1991} based on a wave-kinetic equation and assuming scale-nonlocal interactions between waves and ZFs. {Due to  nonlocality, the evolution of the small scales is described by a linear diffusion equation in $\mathbf{k}$-space with an anisotropic diffusion tensor. This tensor turns out to be degenerate so that diffusion of the small-scale spectral energy density occurs along one-dimensional curves, $\kappa\,k_x - \omega_\mathbf{k} =$ constant. A summary of this work is available at in~\cite{NonlocalDWarxiv2010}.
This mechanism was confirmed numerically by DNS of the forced/dissipated CHM model with an instability forcing (i.e. $\lambda_{\bf k} \ne 0$ in Eq.~\eqref{eq-driftdispersion}) in~\cite{ConnaughtonEPL2011}}.

Another mechanism to suppress small-scale turbulence via shearing of small vortices by the ZF was suggested in~\cite{Biglari1990}. Their argument is typical for 2D incompressible fluids and thus describes the strongly turbulent, vortex-dominated regime rather than WT. The conditions for realising the wave or vortex turbulence regimes in real fusion devices and the respective saturated ZF strengths are	 discussed in~\cite{ConnaughtonEPL2011}. 

\begin{figure}
\begin{center}
\includegraphics[width=\columnwidth]{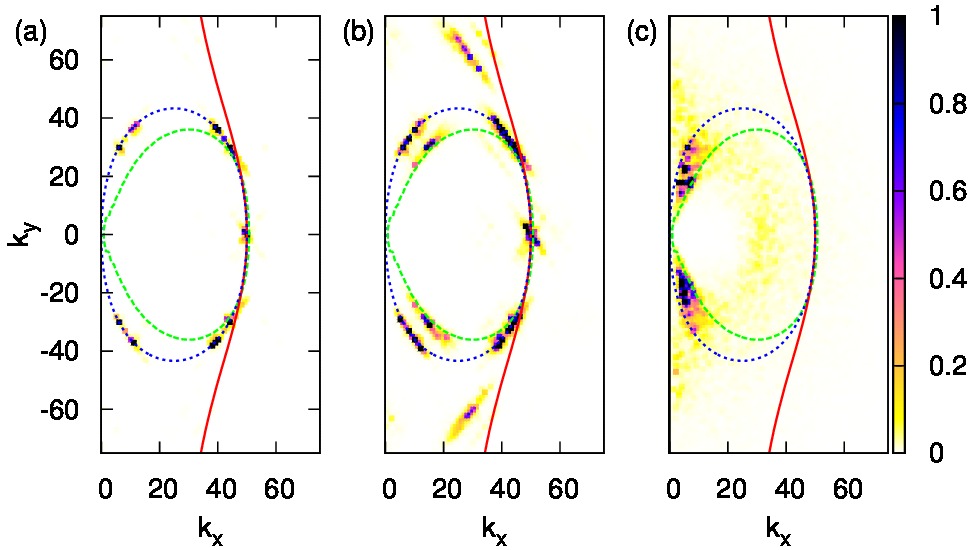}
\end{center}
\caption{ Evolution of the energy spectrum out of a random seed in the 2D $\bf k$-space for narrowband forcing. { The solid curve is $\kappa\,k_x-\omega_\mathbf{k}=$ constant, the dotted line is the resonant manifold of the carrier wave and the dashed line is the curve of constant zonostrophy (see \cite{ConnaughtonEPL2011}).}}
\label{2d-kspace}
\end{figure}

Self-regulation of the drift/Rossby turbulence and ZF system in the CHM model, incorporating both wave and the vortex regimes, was given in \cite{Nazarenko2011} and thus will only briefly be discussed here.  Numerical investigations ~\citep{ConnaughtonEPL2011} traced the first stage
of ZF growth to the MI. This is particularly evident for narrow band forcing, where $\lambda_{\bf k} \ne 0$ for one mode only, ${\bf k} = {\bf k}_f$. Figure~\ref{2d-kspace} shows a { Fourier-space view of} the development of turbulence from a random initial condition. The zonal modes emerge in a way similar to section \ref{sec-MI}. Spectral maxima appear in $\bf k$-space at wave-numbers close to the zonal axis.  Subsequent suppression of the spectrum at the primary instability (meridional) scales results in complete blocking of the instability forcing and saturation of the ZF. Two types of scaling of the saturated ZF magnitude $U$ with the primary instability growth rate $\gamma_{max}$ were predicted theoretically for the WT and the vortex regimes:
$U \propto \sqrt{\gamma_{max}}$  and  $U \propto \gamma_{max}$ respectively.
These scalings were clearly observed numerically.  

\begin{figure}
\begin{center}
\includegraphics[width=\columnwidth]{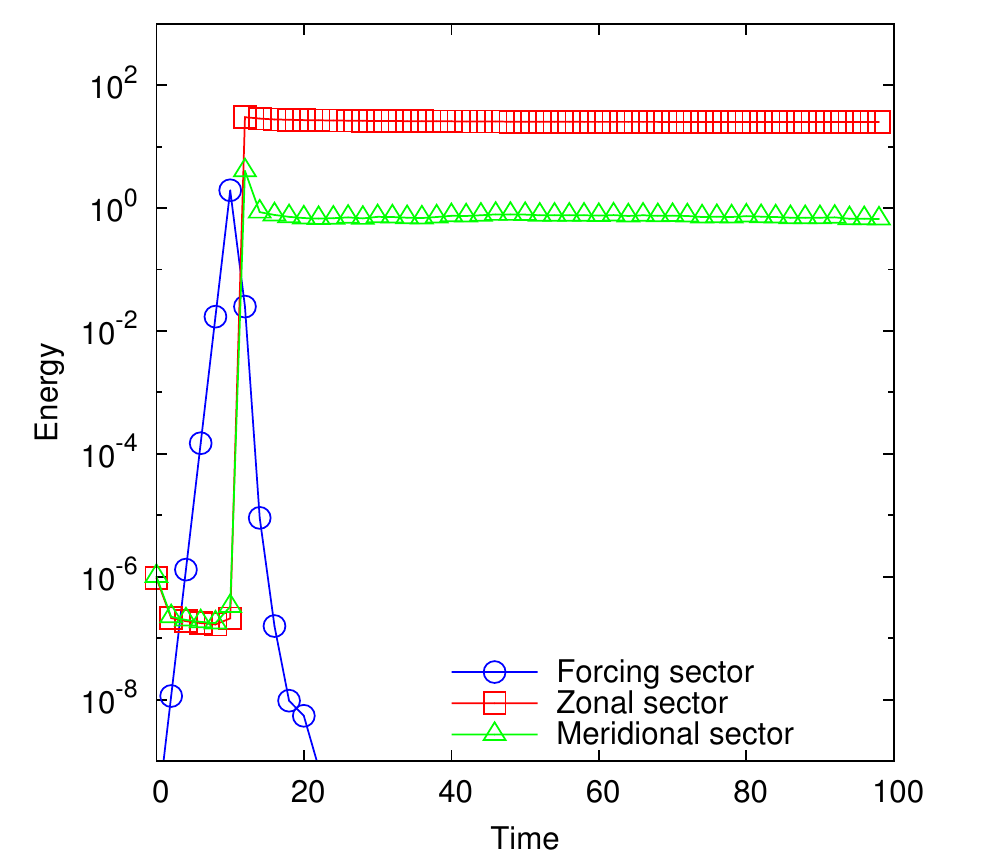}
\end{center}
\caption{ Evolution of the energy contained in the forcing mode, zonal and meridional sectors for narrowband forcing.}
\label{narrowforcing}
\end{figure}

Figure \ref{narrowforcing} shows the evolution of the energy contained in the forcing mode ${\bf k}_f$, zonal sector 
$|k_y|>|k_x|$ and meridional sector $|k_x|>|k_y|$. Initially the forcing mode grows exponentially with the growth rate of the 
primary instability until a significant zonal component forms. This is followed by almost complete suppression  of the forcing mode as well as all
the modes in the meridional sector and saturation of the energy in the zonal sector.

\subsection{Feedback loop in EHM}

A study with the idealised single-mode forcing in the EHM model reveals a qualitatively similar behaviour for the spectrum diffusion in $k$-space and for the suppression of the forcing mode.  It does however, display slightly better zonation properties in that the zonal sector saturates a slightly higher level.

\subsection{Feedback loop in HW and EHW }
The drift turbulence - ZF feedback loop in HW and EHW was first studied by~\cite{Numata2007}. 
ZF generation was found for EHW but not for HW leading the authors to conclude that there is no ZF generation and feedback mechanism in  HW and, therefore, that the extension of the model is crucial. However, this is true only for relatively small coupling parameter values, $\alpha \lesssim 1$.
Recently~\cite{pushkarev2013zonal} studied the HW model with a range of $\alpha$.
They showed that for large $\alpha$'s the HW system does generate ZF's
which suppress the primary instability and saturate. This is natural since HW tends to CHM for large $\alpha$. However, to observe the WT-ZF feedback 
mechanism in CHM requires an ad-hoc forcing term, whereas in the HW system instability is present naturally, even though it is weak for large $\alpha$.

\section{Summary}

We have reviewed the theory of modulational instability for Rossby/drift waves in basic GFD/plasma models: unforced/forced CHM/EHM and HW/EHW equations. We described the linear instability theory using the four-mode truncation (4MT) and emphasised the role of the degree of nonlinearity, $M$, the deformation radius $\rho$ and resonant wave interactions in the weakly nonlinear regime. 
A bifurcation of the most unstable modulation from zonal to off-zonal occurs when
$M$ falls below a critical value: 0.53 for CHM with $\rho=\infty$ and a lower ($\rho$-dependent) value for EHM. This bifurcation affects the late-stage nonlinear where numerics show the generation of off-zonal random jets in the weakly nonlinear case ($M=0.1$). This may play a role in the formation of inclined quasi-zonal jets in both GFD and plasmas.

By comparing with DNS of the CHM and EHM models, we saw that the  4MT works well for small nonlinearities, $M$. It also works well for the initial evolution in the strongly nonlinear case, $M \gtrsim 1$ and accurately predicts the value of $M$ for which a transition between the (weakly nonlinear) oscillatory to  the (strongly nonlinear) saturated long time behavior is observed.
The nonlinear evolution for small $M$ is dominated by wave dynamics whereas for large
$M$ the nonlinear evolution leads to rolling up of the carrier wave vorticity into Karman-like vortex streets. Such vortices behave very differently from waves and it it precisely at the moment of roll-up that the full system's evolution strongly departs from the prediction of the 4MT. After the roll-up, the full system enters into a saturated quasi-stable state which persists for a relatively long time before eventually decaying due to small scale dissipation.  For the zonal dynamics,  we observe the formation of narrow  jets which are more stable than would be expected from the Rayleigh-Kuo criterion  because their 2D structure consists of stable vortex streets. Such narrow jets are very effective transport barriers in both plasmas and GFD. If $M$  is small and 
roll-up does not occur (or is delayed) the full system may follow its 4MT counterpart for much longer, exhibiting nonlinear oscillations as in the 4MT.

We also compared the extended CHM and HW models with the original versions. The extended models generally form zonal flows more easily. The HW model has a different MI mechanism to the CHM model. It occurs as a beating of the carrier wave with its satellites pumping a zonal mode. The resulting ZF initially grows exponentially at a double the maximum growth rate of the primary instability. Late stages in HW show suppression of the primary instability and saturation of the ZF for large values of the coupling parameter, $\alpha$. No such suppression occurs for lower values of $\alpha$.   For EHW, suppression occurs for lower values
of $\alpha$.  This process of ZF generation followed by suppression of small scale turbulence is also evident in simpler forced-dissipated CHM model. The three stages of evolution - generation of  meridional waves by a primary instability, MI of the meridional waves and generation of ZF and suppression of the  small-scale turbulence by the ZF shear - comprise, in a nutshell, the LH transition mechanism.

\bibliography{bibliography}
\end{document}